\documentclass[reprint,pra,aps,superscriptaddress,showpacs,final,floatfix,twocolumn]{revtex4-2}

\usepackage{graphicx}
\usepackage{dcolumn}
\usepackage[mathlines]{lineno}
\usepackage{xcolor}
\usepackage[T1]{fontenc}
\usepackage[utf8]{inputenc}
\usepackage{dcolumn}
\usepackage{pifont}
\usepackage{amsmath}
\usepackage{amssymb}
\usepackage{hyperref}
\usepackage{subfigure}
\usepackage{footnote}
\usepackage{stmaryrd}
\usepackage{bm}
\usepackage{xcolor}
\usepackage{enumitem}
\makeatletter

\usepackage[normalem]{ulem}

\begin{document}

\title{Spikes in Poissonian quantum trajectories}

\author{Alan Sherry} 
\email{alan.sherry@icts.res.in}
\affiliation{International Centre for Theoretical Sciences, Tata Institute of Fundamental Research, Bangalore 560089, India}
\author{Cedric Bernardin}
\email{sedric.bernardin@gmail.com}
\affiliation{Faculty of Mathematics, National Research University Higher School of Economics, 6 Usacheva, Moscow, 119048, Russia.}
\author{Abhishek Dhar}
\email{abhishek.dhar@icts.res.in}
\affiliation{International Centre for Theoretical Sciences, Tata Institute of Fundamental Research, Bangalore 560089, India}
\author{Aritra Kundu} 
\email{aritra.kundu@uni.lu}
\affiliation{Department of Physics and Materials Science, University of Luxembourg, L-1511 Luxembourg}
\author{Raphael Chetrite}
\email{raphael.chetrite@univ-cotedazur.fr}
\affiliation{Institut de Physique de Nice (INPHYNI), Universit\'{e} C\^{o}te d’Azur, CNRS, 17 rue Julien Laupr\^{e}tre, Nice, 06200, France}

\begin{abstract}
We consider the dynamics of a continuously monitored qubit in the limit of strong measurement rate where the quantum trajectory is described by a stochastic master equation with Poisson noise. Such limits are expected to give rise to quantum jumps between the pointer states associated with the non-demolition measurement. A surprising discovery in earlier work [Tilloy \emph{et al.}, Phys. Rev. A {\bf 92}, 052111 (2015)] on quantum trajectories with Brownian noise was the phenomena of spikes observed in between the quantum jumps.  Here, we show that spikes are observed also for Poisson noise. We consider three cases where the non-demolition is broken by adding, to the basic strong measurement dynamics, either unitary evolution or thermal noise or additional measurements. We present a complete analysis of the spike and jump statistics for all three cases using the fact that the dynamics effectively corresponds to that of stochastic resetting. We provide numerical results to support our analytic results.  
\end{abstract}

\maketitle

\tableofcontents

\section{Introduction }

In the last decade, the flourishing interest in feedback control
of quantum systems has led to a renaissance of the theory of quantum
measurements through continuous measurements. The cornerstone of 
continuous quantum measurements is the stochastic master equation
(SME), also called the quantum trajectories equation, or the Belavkin-{Barchielli}
equation. After an initial period of development in the spontaneous
collapse community \cite{Bohm,Pearle,Pearle-86,Diosi}, SME was really
born in the eighties in quantum optics and quantum filtering communities
\cite{Gisin,Barchielli-06,Alicki,Diosi88,Belavkin89,Belavkin90,Barchielli,Belavkin92,Belavkin,Holland,Gardiner92,Dalibard92,Wiseman-1,Garraway94}. They are now part of standard textbooks~\cite{Gardiner,Wiseman,Haroche,Breuer,Carmichael,Jacobs-Livre},
and have been observed experimentally \cite{Guerlin,Gleyzes,Sayrin,Murch2013,Murch2013-2,Weber}. In the context of the so-called Quantum Non-Demolition (QND) setup it is assumed that measurement operators,  and possible Hamiltonians, are all diagonalisable in the same orthonormal basis, called the pointer basis. 
  One of the most striking results in this case is that the large time-limiting form of the density matrix is in accordance with the usual Born rule. This has been  observed experimentally~\cite{Guerlin,Gleyzes,Sayrin} and  proven theoretically in \cite{Adler,Van-Handel,Maassen,Bauer2011,Benoist-2014}.

An interesting situation arises when one considers the non-QND setup (see Sec.~\ref{sec:general-setup}), where a lot of attention has recently been given to the case of large rate measurement limit, which is also called the strong collapse limit (or weak non-QND limit) of non-QND quantum trajectories. We then expect to see the emergence of quantum jumps between pointer states and  this has been   observed experimentally~\cite{Bergquist,Nagourney,Sauter}. It was recently theoretically explained in~\cite{Bauer15} for SME with Gaussian noise in the Collapse-Thermal and Collapse-Unitary setups. \cite{Benoist} generalised \cite{Bauer15} for SME with Gaussian noise, in more general setups, particularly the Collapse-Measurement case.  \cite{Ballesteros} also discusses it for SME with Poisson noise in the Collapse-Unitary setup.  Note that the quantum Zeno effect~\cite{Degasperis,Misra,Peres,Itano,Teuscher,Layden}  argues that continuously observing a quantum state leads to a zero probability of evolving away from a given state, thus freezing the evolution of the system and suppressing the quantum jumps. However, the Zeno effect can be avoided if one considers the appropriate limit of system probe interactions.

A surprising discovery was the observation of sharp scale-invariant fluctuations that invariably decorate the jump process (see Fig.~\ref{fig:traj-rabi}c, \ref{fig:traj-thermal}c and \ref{fig:traj-coll}c), in the limit where the measurement rate is very large. These seemingly instantaneous excursions have been referred to as \textit{quantum spikes}.  These fluctuations around the dominant jump process seem to be visible in the early numerical work on the subject \cite{Gisin92,Mabuchi,Cresser06,Colin12}. A heuristic description was provided in \cite{Cresser06,Colin12},  while their first analytical treatment  was done recently by Bauer-Bernard-Tilloy~\cite{Bauer,Tilloy,Bauer2018}, which motivated some other works~\cite{Kolb,BCCNP2023}. Previous work has been exclusively concerned with  Gaussian noise SMEs and explains these phenomena by relating it to the fact that the convergence to the jump process between pointer states is weak. However, this is expected to be valid for more general SMEs and we quote Tilloy \cite{Tilloy2019}: 
\emph{``A first possible generalization is to go from
a continuous collapse model to a discrete one, and replace the diffusive
equations we had with jump ones......However, while spikes are numerically
present as well in this context and seem to have the exact same power
law statistics, no proof is known even for the qubit case.''}
The main goal of the current paper is to demonstrate the phenomena of quantum spikes for the case of Poisson noise SMEs and to study their statistical properties.

We limit our analytical study to particular cases of a qubit system, for  the three setups which we refer to as the   Collapse-Unitary setup, the Collapse-Thermal setup, and the Collapse-Measurement setup. The restriction of our study, apart from the fact that it concerns only a qubit, is that its analytical resolution is based on the fact that the resulting dynamics is in fact a $1$-dimensional  piecewise resetting Markov dynamics \cite{Evans11,Evans11-2,Evans20,Pal16}, {and explicit Laplace transforms lead to the complete statistics of the associated quantum spikes.} In Sec.~\ref{sec:conjecture} we discuss our results in the context of more general piecewise deterministic Markov processes.

The plan of the paper is as follows: in Sec.~\ref{sec:general-setup},  we discuss the general set-up and then in Sec.~\ref{sec:precisesetup}, explain the three specific models that we study here. In Sec.~\ref{sec:mainresults}, we summarize the main results and present comparisons of various analytic expressions with numerical data. The mathematical derivations are presented in Sec.~\ref{sec:derivations}. In Sec.~\ref{sec:conjecture}, we enlarge our discussion for {more abstract mathematical models} and state a conjecture about the spikes statistics for the latter. We conclude the paper with a conclusion section, see Sec.~\ref{sec:conclusion}, where we also discuss the relevance of spikes in real experimental studies.

\section{General setup: stochastic master equation (SME) of $d-$dimensional systems with Poisson noise}
\label{sec:general-setup}

We consider here SMEs with Poisson white noise (discarding the possible
Gaussian part of the noise), in finite $d$-dimensional Hilbert space,
describing the real-time evolution of a quantum system with (Hermitian) density
matrix $\rho$ ($\rho\geq 0, {\rm {Tr}} (\rho)=1$)
with free evolution given by a Hermitian
Hamiltonian $H$, thermal evolution given by a Lindbladian $L^{th}$~\cite{Gorini,Lindblad}, and simultaneous  independent continuous measurement of $p$ operators  $\left\{ N_{k} \in \mathcal{M}_{d} (\mathbb{C})\right\}_{k=1, \ldots,p}$, where $\mathcal{M}_{d} (\mathbb C)$ denotes the set of square complex matrices of size $d$, and with  
with rate of measurement $\gamma=\{ \gamma_k\}_{k=1, \ldots, p}$. The evolution equation of the density matrix is given by the piecewise deterministic Markov process (see for example relations (6.224-6.227) of \cite{Breuer})
\begin{equation}
\label{eq:EMS-3}
\begin{split}
&{\dot \rho}_{t}^{\gamma}  =-i\left[H, \rho_{t}^{\gamma}\right] + L^{th}\left[ \rho_{t}^\gamma \right]\\
&+\sum_{k=1}^{p}\left\{ \gamma_{k}L_{N_{k}}\left[ \rho_{t}^{\gamma}\right]+M_{N_{k}}\left[ \rho_{t}^{\gamma}\right]\left[ \dot{\mathcal{N}}_{t}^{k, \gamma} -\mathbb{E}\left( \dot{\mathcal{N}}_{t}^{k, \gamma}\vert \rho_{t}^{\gamma}\right)\right]\right\} \ .
\end{split}
\end{equation}
In the previous equation: 
\begin{itemize}[label=$\ast$]
\item The  (classical) \textit{signal measurements} $\{ \mathcal{N}_{t}^{k, \gamma }\} _{k=1, \ldots p}$
are given by non-autonomous Poissonian processes ($d\mathcal{N}_{t}^{k, \gamma} \in\{0,1\}$ and  $d\mathcal{N}_{t}^{k, \gamma}d\mathcal{N}_{t}^{\ell, \gamma}=d\mathcal{N}_{t}^{k, \gamma }\delta_{k \ell}$), with statistics 
\begin{equation}
\label{eq:Poi-1}
\mathbb{E}\left( d\mathcal{N}_{t}^{k, \gamma}\vert \rho_{t}^{\gamma}\right)=\gamma_{k}\eta_{k}{\rm{Tr}} \left(N_{k}\rho_{t}^\gamma N_{k}^{\dagger}\right)dt  \ .
\end{equation}
We will refer to the signals $d \mathcal{N}_t^{k,\gamma}=1$ as "clicks".
Note that, in all the terms,  $\rho_{t}^{\gamma}$ in the previous equations 
must be understood as the density matrix taken just before an eventual
jump happening at time $t$.

\item The \textit{measurement operators} $N_{k}$ are arbitrary matrices (not necessarily commuting, not necessarily Hermitian) and characterise each independent detector.  The parameters $\eta_{k}\in\left[0,1\right]$ are the associated detector efficiencies: $\eta_{k}=1$ (resp. $\eta_k=0$) corresponds to a perfect efficient detector (resp. an unread measurement).

\item The \textit{Lindbladian} {\footnote{completely positive trace
preserving map from $M_{d}(\mathbb{C})$ to $M_{d} (\mathbb{C})$.}} operator $L_{N}$, associated to the measurement operator $N$,  encapsulates the back-effect of the $N$-measurement on the average evolution. It is given by \cite{Gorini,Lindblad}
\begin{equation}
\label{eq:Lin}
L_{N}\left[\rho\right] =  N\rho N^{\dagger}-\tfrac{1}{2}\, N^{\dagger}N\rho-\tfrac{1}{2}\, \rho N^{\dagger}N \ . 
\end{equation}

\item The {\textit{stochastic innovation} }$M_{N}\left[\rho\right]$ is the non-linear operator defined by
\begin{equation}
M_{N}\left[\rho\right] = \tfrac{N\rho N^{\dagger}}{{\rm{Tr}}\left(N\rho N^{\dagger}\right)}-\rho \ .
\label{eq:M}
\end{equation}
\end{itemize}

From a theoretical viewpoint, the simplest way to derive the SME Eq.~\eqref{eq:EMS-3} from ideal physical systems is as the limit of infinitely frequent indirect measurements  (with the system-ancilla interaction becoming infinitely strong in the limit), see for example \cite{Breuer,Pellegrinni,Jacobs-Livre, Wiseman,Snizhko,DCD2023}. The non-linearity in Eq.~\eqref{eq:EMS-3} is a remnant of the non-linear updating of all type of Bayesian measurements. However, this non-linearity is in fact the effect of the fixed normalisation ${\rm {Tr}}(\rho_{t}^\gamma)=1$, since, without this normalisation,  Eq.~\eqref{eq:EMS-3} can be formulated as a  linear equation  \cite{Gardiner,Wiseman,Haroche,Breuer,Carmichael,Jacobs-Livre}.

Note that in practice, for a real experiment, the knowledge about the quantum system is only stored in the Poisson signal process $\{ \mathcal{N}_{t}^{k, \gamma}\} _{k\in 1, \ldots, p}$. The density matrix $\rho_{t}^\gamma$ itself, is not directly observable, and is only reconstructed from the Poisson signal process.

\subsection{Large time limit of the SME Eq.~\eqref{eq:EMS-3}: to collapse or not }

The first crucial question concerning the SME Eq.~\eqref{eq:EMS-3} is the understanding of its large-time behaviour. Let us assume for the moment that we do not have any thermal bath $L^{th}=0$. The \textit{Quantum Non-Demolition (QND)} hypothesis~\cite{Braginsky75,Thorne78,Unruh78,Braginsky,Caves80,Milburn83,Haroche,Guerlin,Wiseman} is the fact that there exists an orthonormal basis $\{ |e_{i}\rangle\} _{i=1, \ldots, d }$, called the \textit{pointer basis}~\cite{Zurek}, such that all the measurement operators $\{ N_{k}\} _{k=1, \ldots, p}$, and the Hamiltonian $H$, are diagonal in this basis.  Under this hypothesis (and a non-degeneracy condition that we do not discuss here, see \cite{Benoist-2014}), the large time limit of the density matrix in Eq.~\eqref{eq:EMS-3} is given by the famous collapse-absorbing formula
\begin{equation}
\label{eq:Collapse}
\lim_{t\rightarrow\infty}\rho_{t}^{\gamma}=\vert e_{i}\rangle \langle e_{i}\vert \textrm{ with probability } \langle e_{i},\rho_{0}^{\gamma}e_{i}\rangle \  .
\end{equation}
This corresponds to the Born rule for measurement outcomes under projective measurements of any of the $N_k$ at the initial time.
Related results have been observed experimentally~\cite{Guerlin,Gleyzes,Sayrin} and are proven theoretically in \cite{Adler,Van-Handel,Maassen,Bauer2011,Benoist-2014}.

\bigskip

However, there are many ways to break the QND hypothesis, and we discuss three different ones:
\begin{enumerate}
\item \textit{Collapse-Unitary} setup: We may allow that the Hamiltonian $H$ is not diagonal in the pointer basis $\{ \vert e_{i}\rangle\} _{i=1, \ldots,d}$ related to the measurement operators $\{N_k\}_{k=1, \ldots, p}$. It results then in a competition between collapse in the pointer basis and a unitary evolution (non-collapsing) generated by $H$.

\item  \textit{Collapse-Thermal} setup: We may take into account the thermal interactions with the external world, i.e. $L^{th}\neq0.$ Then it results in a competition between collapse Eq.~\eqref{eq:Collapse}  in the pointer basis,  and,  thermalisation towards a Gibbs state.

\item \textit{Collapse-Measurement}  setup: We may allow that one of the measurement operator, say $N_{p}$, is not diagonal in the pointer basis $\{ \vert e_{i}\rangle \} _{i=1,\ldots,d }$ related to $\{H, \{N_k\}_{k \ne p} \}$. It results then in a competition between collapse Eq.~\eqref{eq:Collapse} in the pointer basis of $\{H, \{N_k\}_{k \ne p} \}$ and collapse in the pointer basis related to $N_p$,  if $N_p$ is diagonalisable. If $N_p$ is not diagonalisable, there is competition between collapse in the pointer basis  related to $\{H, \{N_k\}_{k \ne p} \}$ and the uncollapse dynamics generated by the measurement operator $N_p$. 
\end{enumerate}

In all these three setups, the large time collapse Eq.~\eqref{eq:Collapse} is then a priori broken.

\subsection{Strong collapse Zeno limit in a non-QND setup: Quantum jump phenomenon versus Spikes phenomenon}

In this work, instead of the large time limit we consider the large $\gamma$ strong collapse limit, also called the quasi-Zeno limit. For non-QND quantum trajectories, Eq.~\eqref{eq:EMS-3} leads to a richer phenomenology:

\begin{itemize}[label=$\ast$]

\item {\textit{Quantum jumps}}: The expected quantum jumps definitively arise between the pointer basis. More formally, the density matrix will converge to a pure jump Markov process 
\begin{equation}
\lim_{ \gamma  \rightarrow\infty}\rho_{t}^{\gamma}=\vert X_{t}\rangle \langle X_{t}\vert  ,\label{eq:ConvSaut}
\end{equation}
where $X_{t}\in\left\{ \vert e_{i} \rangle \right\} _{i=1,\ldots,d}$ is a pure jump Markov process with explicit computable rates \cite{Bauer15,Ballesteros,Benoist}. In some cases they can vanish.  For example, the transition rate between pointer states
decrease as $\frac{1}{\vert \gamma \vert}$ for large $\gamma$ in the Collapse-Unitary setup, for a fixed $H$ (this is reminiscent of the Zeno effect).

\item  {\textit{Quantum Spikes}}:  One finds that sharp scale-invariant fluctuations invariably decorate the jump process (see Figs. \ref{fig:traj-rabi}c, \ref{fig:traj-thermal}c and \ref{fig:traj-coll}c below), in the limit where the measurement rate is very large. We call \textit{quantum spikes} these seemingly instantaneous excursions. 
 These spikes become infinitely thin when $\gamma\rightarrow\infty$, i.e. take a time $\frac{1}{\vert \gamma \vert}$ while their heights remain of
order $1$ in the limit. 
As discussed above, such spikes have been observed and explained for the case with Gaussian noise~\cite{Bauer,Tilloy,Bauer2018}. These articles  explain the spiking phenomena as the fact that the convergence in Eq.~\eqref{eq:ConvSaut} is weak. Exporting these methods in the present context seems to be a difficult task since the authors there are using tools specific to Brownian motion. 
\end{itemize}

\section{Precise setup: SME of Qubit for Non-Demolition ``reset'' continuous measurement of $\vert-\rangle \langle -\vert$}
\label{sec:precisesetup}

\label{subsec:Non-Demolition-sharp}

We consider a qubit system with basis $\{ \vert+\rangle ,\vert-\rangle\} $, where continuous collapse measurement competes with a thermal, an other measurement or a unitary dynamics.

Precisely, we consider the collapse measurement of a rank one diagonal (see QND hypothesis) operator
\begin{equation}
\label{eq:operatorN}
N_1=\left(\begin{array}{cc}
0 & \text{0}\\
0 & 1
\end{array}\right)=\vert -\rangle \langle -\vert \ .
\end{equation}

\medskip

The jumps in the SME Eq.~\eqref{eq:EMS-3} due to the term $M_{N_1}[\rho_t^\gamma]\,  \dot {\mathcal N}_t^{1,\gamma}$ are, thanks to Eq.~\eqref{eq:M}, given by
\begin{equation}
\label{eq:resetting}
\rho \rightarrow \rho+M_{N_1}[\rho] = \cfrac{N_1 \rho N_1^{\dagger}}{{\rm{Tr}}\left(N_1 \rho N_1^{\dagger}\right)}=\vert -\rangle \langle -\vert,
\end{equation}
i.e. the result after the jump is always the same matrix, irrespective of the state just before the jump: this is a resetting dynamics  \cite{Evans11,Evans11-2,Evans20,Pal16}. In fact, the crucial property to get resetting dynamics is not the precise form of $N_1$ but the fact that the matrix $N_1$ is of rank one  (''sharp measurement''). Since we are considering a qubit, note also that the only other rank one diagonal measurement operator satisfying QND hypothesis is
\begin{equation}
\label{eq:N_1other}
\left(\begin{array}{cc}
1 & \text{0}\\
0 & 0
\end{array}\right)=\vert +\rangle \langle +\vert \ ,
\end{equation}
which gives exactly the same theory by the exchange $\vert + \rangle \to \vert -\rangle$. Hence our focus will be on $N_1$.
{Note that in Section \ref{sec:conjecture}, we will discuss an  extension to general diagonal $N_1$.}

\medskip

By parameterising the density matrix as usual as
\begin{equation}
\label{eq:parameter1}
\rho_{t}^\gamma=\left(\begin{array}{cc}
q_{t}^{\gamma} & {\overline u}_{t}^{\gamma}\\
u_{t}^{\gamma} & 1-q_{t}^{\gamma}
\end{array}\right)\ ,
\end{equation}
with $(q_t^\gamma,u_t^\gamma) \in \mathbb{R}\times \mathbb{C}$,
we get explicitly in Appendix \ref{app:1} the analytical form of the $N_1$-measurement part (in the SME Eq.~\eqref{eq:EMS-3}) for general diagonal $N_1$.  For  the particular case Eq.~\eqref{eq:operatorN}, it takes the form of a three-dimensional process:
\begin{equation}
\label{eq:sde-2-1-2-1-10}
\begin{cases}
\dot q_{t}^{\gamma}=-q_{t}^{\gamma}\left(\dot{\mathcal N}_{t}^{1,\gamma}-\gamma_1 \eta_1 \left(1-q_{t}^{\gamma}\right)\right) \ , \\
\dot u_{t}^{\gamma}=-\frac{\gamma_1}{2}u_{t}^{\gamma}-u_{t}^{\gamma}\left(\dot{\mathcal N}_{t}^{1,\gamma}-\gamma_1 \eta_1 \left(1-q_{t}^{\gamma}\right)\right) 
\end{cases}
\end{equation}
with
\begin{equation*}
 \mathbb{E}\left( d{\mathcal N}_{t}^{1,\gamma}\vert q_{t}^{\gamma}\right)=\gamma_1 \eta_1 \left(1-q_{t}^{\gamma}\right)dt , \quad 
 d{\mathcal N}_{t}^{1,\gamma}d{\mathcal N}_{t}^{1,\gamma}=d{\mathcal N}_{t}^{1,\gamma} \ .
\end{equation*}

Note that the resetting Eq.~\eqref{eq:resetting} corresponds in the coordinates Eq.~\eqref{eq:parameter1} to the resetting to $(q=0,u=0)$. Moreover the equations in  Eq.~\eqref{eq:sde-2-1-2-1-10} show that  the population $q_{t}^{\gamma}$ is a martingale collapsing to $\{0,1\}$ \cite{Adler,Van-Handel,Maassen,Bauer2011,Benoist-2014,roldan2023martingales}. Moreover, in average,  $u_{t}^{\gamma}$ shrinks trivially to zero at large time, or for strong measurement $\gamma_1 \to \infty$ (decoherence).
We will study analytically three specific cases
where we add either a unitary drive or a thermal bath or a non-commuting measurement.

\subsection{Collapse-Unitary setup: Measurement of Rabi qubit coherent oscillations}

We choose here no extra measurement apart from $N_1$, and
\begin{equation}
\label{eq:Rabi}
L^{th}=0,\quad H=k_{\gamma_1} \sigma_{x}=k_{\gamma_1}\Big(\begin{array}{cc}
0 & 1\\
1 & 0
\end{array}\Big) \ .
\end{equation}

 This Rabi Hamiltonian $H$ and the measurement operator $N_1$ given in Eq.~\eqref{eq:operatorN} do not commute and then, they are not diagonalisable in a common orthonormal basis. Moreover, the ad-hoc dependency in $\gamma_1$ of the Rabi frequency $k_{\gamma_1}$
(which is, in general, proportional to the magnetic field) will stay unfixed for the moment, and will be discussed below in relation to
the quantum Zeno effect. The Hamiltonian contribution is given by 
 \begin{equation}
 -i\left[H,\rho_{t}^{\gamma}\right]  =-ik_{\gamma_1}\left(\begin{array}{cc}
u_{t}^{\gamma}-{\overline u}_{t}^{\gamma} & 1-2q_{t}^{\gamma}\\
2q_{t}^{\gamma}-1 & {\overline u}_{t}^{\gamma}-u_{t}^{\gamma}
\end{array}\right) \ ,
\end{equation}
and the SME Eq.~\eqref{eq:EMS-3} takes then the form of a $3$-dimensional piecewise deterministic Markov process 
\begin{equation}
\label{eq:sde-2}
\begin{cases}
\dot q_{t}^{\gamma}=-ik_{\gamma_1}\left(u_{t}^{\gamma}-{\overline u}_t^{\gamma}\right)-q_{t}^{\gamma}\left(\dot {\mathcal N}_{t}^{1,\gamma} -\gamma_1\eta_1\left(1-q_{t}^{\gamma}\right)\right)\\
\dot u_{t}^{\gamma}=-ik_{\gamma_1}\left(2q_{t}^{\gamma}-1\right)-\frac{\gamma_1}{2}u_{t}^{\gamma}-u_{t}^{\gamma}\left(\dot {\mathcal N}_{t}^{1,\gamma} -\gamma_1 \eta_1 \left(1-q_{t}^{\gamma}\right)\right) 
\end{cases}
\end{equation}
with 
\begin{equation*}
 \mathbb{E}\left(d{\mathcal N}_{t}^{1,\gamma}\vert q_{t}^{\gamma}\right)=\gamma_1 \eta_1 \left(1-q_{t}^{\gamma}\right)dt, \quad 
 d{\mathcal N}_{t}^{1,\gamma}d{\mathcal N}_{t}^{1,\gamma}=d{\mathcal N}_{t}^{1, \gamma} \ .
\end{equation*}

Note that contrary to the Collapse-Thermal setup and the Collapse-Collapse setup,  discussed in the following two sections,  the real population $q_{t}^{\gamma}$ is not autonomous: it depends of the complex coherence $u_{t}^{\gamma}$. Finally, when the rate of measurement vanishes (i.e. $\gamma_1=0)$, this is the usual equation for Rabi Oscillation \cite{Feynman,Le-Bellac}: precession of a spin one-half in an external magnetic field.

Parametrise now the density matrix in Bloch coordinates: 
\begin{equation}
\label{eq:QIS-1}
\rho_{t}^\gamma=\frac{1}{2}\left(\begin{array}{cc}
1+r_{t}^{\gamma}\left(\cos\theta_t^{\gamma} \right) & r_{t}^{\gamma}\left(\sin\theta_t^{\gamma} \right)\exp\left(-i\varphi_t^{\gamma} \right)\\
r_{t}^{\gamma}\left(\sin\theta_t^{\gamma} \right)\exp\left(i\varphi_t^{\gamma} \right) & 1-r_{t}^{\gamma}\left(\cos\theta_t^{\gamma} \right)
\end{array}\right)\ ,
\end{equation}
with {\footnote{This parameterisation is quite unusual but it is convenient for our purpose.}}
\begin{equation}
0\leq r_{t}^{\gamma}\leq1, \quad \theta_t^{\gamma} \in\left]-\pi,\pi\right], \quad \varphi_t^{\gamma} \in\left[0,\pi\right] \ .
\end{equation}
We show in Appendix \ref{sec:Appendix-:-Prove}, Eq.~\eqref{eq:Eqn-r},  that in the totally efficient case $\eta_1=1$, Bloch sphere of pure states $r_{t}^{\gamma}=1$ is stable under the evolution Eq.~\eqref{eq:sde-2}. We will now restrict our study to this totally efficient case and consider only pure state dynamics. We show then in Appendix \ref{sec:Appendix-:-Prove} that the equation Eq.~\eqref{eq:sde-2} takes then the form of the $2$-dimensional piecewise deterministic Markov process Eq.~\eqref{eq:sde-angle-1-1} for $\left(\theta_t^{\gamma} ,\varphi_t^{\gamma} \right)$:
\begin{equation}
\label{eq:sde-angle-1-1-1}
\begin{cases}
\dot\theta_t^{\gamma} =\left(-2k_{\gamma_1}\sin\left(\varphi_t^{\gamma} \right)-\frac{\gamma_1}{2}\sin\left(\theta_t^{\gamma} \right)\right)+\left(\pi-\theta_t^{\gamma} \right)\dot{\tilde{\mathcal N}}_{t}^{1,\gamma} \ , \\
\dot\varphi_t^{\gamma} =-2k_{\gamma_1}\frac{\cos\left(\theta_t^{\gamma} \right)}{\sin\left(\theta_t^{\gamma} \right)}\cos\left(\varphi_t^{\gamma} \right)
\end{cases}
\end{equation}
with 
\begin{equation*}
\mathbb{E}\left( d{\tilde{\mathcal N}}_{t}^{1,\gamma}\vert \theta_t^{\gamma} \right)=\gamma_1 \left(\sin\tfrac{\theta_t^{\gamma} }{2}\right)^{2}dt, \quad 
d{\tilde{\mathcal N}}_{t}^{1,\gamma}d{\tilde{\mathcal N}}_{t}^{1,\gamma}=d{\tilde{\mathcal N}}_{t}^{1,\gamma}.
\end{equation*}

Note that the Poisson noise part effect is a resetting {\cite{Evans11,Evans11-2,Evans20,Pal16}} in the south pole of the Bloch ball $\theta=\pi$ (see Eq.~\eqref{eq:resetting}). 
 
It is clear from the second equation of Eq.~\eqref{eq:sde-angle-1-1-1}
that the plane $\varphi_{t}=\frac{\pi}{2}$ is stable, and then by
restricting the dynamics to this plane,  we see that $\theta_t^{\gamma}$ is the $1$-dimensional piecewise deterministic Markov process given by 
\begin{equation}
\label{eq:sde-angle-1-1-1-1}
\dot \theta_t^{\gamma} =\big(-2k_{\gamma_1}-\tfrac{\gamma_1}{2}\sin\left(\theta_t^{\gamma} \right)\big)+\big(\pi-\theta_t^{\gamma} \big)\, \dot{\tilde{\mathcal N}}_{t}^{1,\gamma} \ .
\end{equation}

This equation was first studied in \cite{Snizhko} and later in \cite{DCD2023} and will be the starting point of our analytical study in the Collapse-Unitary setup, see Section \ref{sec:an-CUsetup}. Again, when $\gamma_1=0,$ this is the usual equation for Rabi oscillations \cite{Le-Bellac,Feynman}, i.e. $\theta_{t}^0=-2k_{0}t$, which implies for the density matrix Eq.~\eqref{eq:QIS-1}
\begin{equation}
\label{eq:QIS-1-1}
\rho_{t}^0=\tfrac{1}{2}\left(\begin{array}{cc}
1+\cos\left(-2k_{0}t\right) & i\sin\left(2k_{0}t\right)\\
-i\sin\left(2k_{0}t\right) & 1-\cos\left(-2k_{0}t\right)
\end{array}\right) \ .
\end{equation}

Note finally that \cite{DCD2023,Snizhko} contains a different derivation of Eq.~\eqref{eq:sde-angle-1-1-1-1}, by starting for a model of indirect measurement \cite{Breuer,Jacobs-Livre} with $2$-dimensional probes.

\begin{figure*}
\includegraphics[width=18cm]{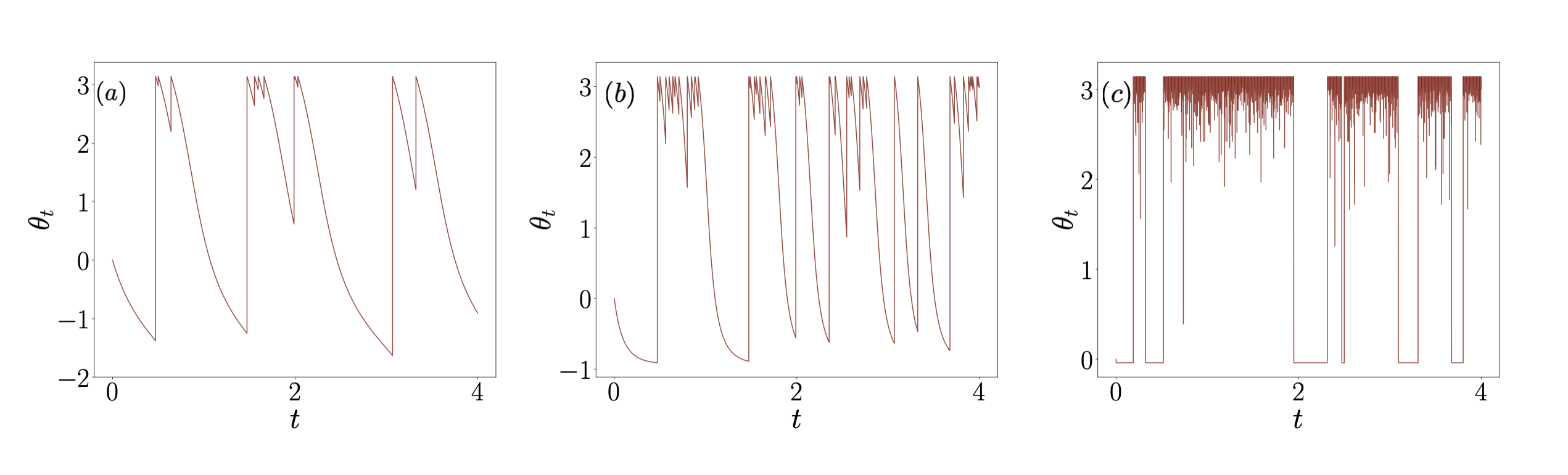}
\caption{Collapse-Unitary setup: 
Typical trajectories generated by Eq.~\eqref{eq:sde-angle-1-1-1-1} for increasing values of $\gamma_1=7,25,10000$ for (a),(b),(c) respectively, with the Rabi frequency tuned as $k_{\gamma_1} = \sqrt{\omega \gamma_1}$. 
In (a) we see time segments with a downward deterministic flow from $\pi$ and then an upward vertical instantaneous reset to $\pi$ --- this constitutes a pre-spike. With increasing $\gamma_1$, these pre-spikes either form a jump to $\theta \approx 0$ or, more often, develop into spikes. Thus  we finally see in (c) a structure consisting of jumps (flows from $\pi$ to $\theta \approx 0$ or instantaneous resets from $\theta \approx 0$ to $\pi$), interspersed with spikes emerging from the state $\theta =\pi$. After a jump to $\theta \approx 0$, the quantum trajectory remains here for a time $\sim 1/(4 \omega)$ during which no spikes occur. In the simulations we took $\omega=1$ and a time step $dt=10^{-5}$ for iterating the Poisson process.}
\label{fig:traj-rabi}
 \end{figure*}

In Fig.~\ref{fig:traj-rabi} we show plots of the trajectory of the process for different values of $\gamma_1$ and the choice  $k_{\gamma_1}=\sqrt{\omega \gamma_1}$ (a negative sign will not change the results). This scaling with $\gamma_1$ is chosen in order to see the spikes (see Appendix \ref{app:Rabi-scaling} for justification).

\begin{figure*}
    \centering
\includegraphics[width=18cm]{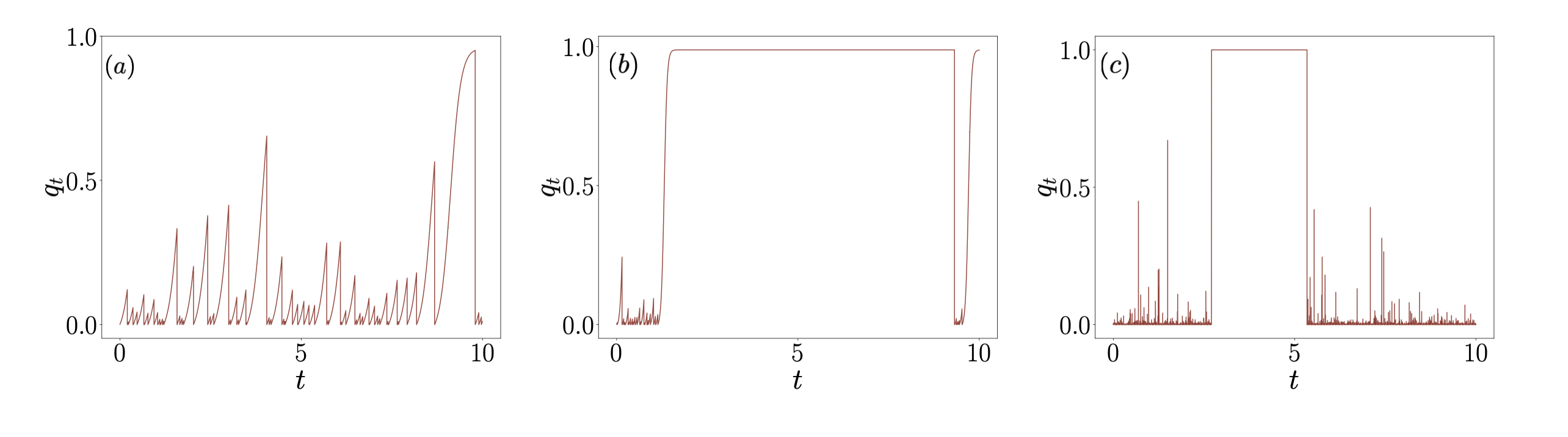}
    \caption{Collapse-Thermal setup: Typical trajectories generated by Eq.~\eqref{eq:EMS-3-1-2-1-1-1-2} for different values of $\gamma_1=7,25,10000$ for (a),(b),(c) respectively. Here we  see that in (a), the trajectory has time segments with pre-spikes which consist of a upward deterministic flow from $q=0$, and a downward reset to the state $q=0$. In the large $\gamma_1$ limit in (c), the pre-spikes develop into sharp spikes and we  again see an effective Poisson jump process between the pointer states at $q=0$ and $q=1$, decorated by spikes emerging from the lower branch.    
    The  parameters here are $W_\mp = W_\pm = 0.3$, $\eta_1 =1$, $dt = 10^{-5}$ }  
    \label{fig:traj-thermal}
\end{figure*}

\begin{figure*}
    \centering
    \includegraphics[width=18cm]{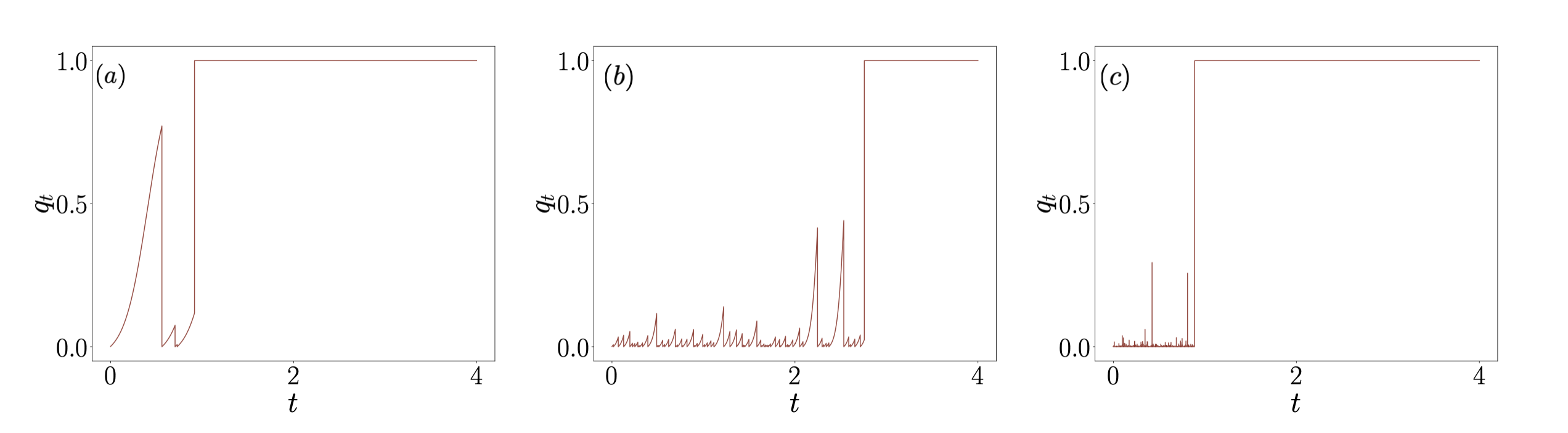}
    \caption{Collapse-Measurement setup: Typical trajectories from Eq.~\eqref{eq:sde-2-2-1-1} for increasing values of $\gamma_1=7,25,10000$ for (a),(b),(c) respectively. In this case the pre-spikes have the same form as in Fig.~\eqref{fig:traj-thermal} and again we see spikes emerging in the limit of large $\gamma_1$. However, in this case, the spikes are transient, since one has a vanishing transition rate from $q=1$ to $q=0$.     
    The parameters are $\gamma_2 = 1.0, \eta_1 = 1.0,\eta_2 = 0.7, dt = 10^{-5}$.} 
   
    \label{fig:traj-coll}
\end{figure*}

\subsection{Collapse-Thermal setup: Measurement of thermal fluctuations}

We choose here $H=0$, no extra measurement apart from $N_1$ and the thermal Lindbladian in a weak coupling form \cite{Davies76}: 
\begin{equation*}
\begin{split}
L^{th}\left[\rho\right] &=  M_{+}\rho M_{+}^{\dagger}-\frac{1}{2} M_{+}^{\dagger}M_{+}\rho-\frac{1}{2} \rho M_{+}^{\dagger}M_{+}\\
&+M_{-}\rho M_{-}^{\dagger} -\frac{1}{2}M_{-}^{\dagger}M_{-}\rho-\frac{1}{2} \rho M_{-}^{\dagger}M_{-} \ , 
\end{split}
\end{equation*}
with 
\begin{equation*}
M_{+}=\sqrt{W_{-,+}}\left(\begin{array}{cc}
0 & 1\\
0 & 0
\end{array}\right), \quad 
M_{-}=\sqrt{W_{+,-}}\left(\begin{array}{cc}
0 & 0\\
1 & 0
\end{array}\right), \ 
\end{equation*} 
where $W_{-,+}, W_{+,-}$ are positive coefficients characterizing the thermal bath.  
We use the parameterisation given in Eq.~\eqref{eq:parameter1}. The components $q_{t}^{\gamma}$ of the SME  Eq.~\eqref{eq:EMS-3} evolve then autonomously as a $1$-dimensional piecewise
deterministic Markov process defined by
\begin{equation}
\label{eq:EMS-3-1-2-1-1-1-2}
\begin{split}
{\dot q}_{t}^{\gamma} & =\left(W_{+,-}+W_{-,+}\right)\left(\tfrac{W_{-,+}}{W_{+,-}+W_{-,+}}-q_{t}^{\gamma}\right) \\
&-q_{t}^{\gamma}\left({\dot{\mathcal N}}_{t}^{1,\gamma}-\gamma_1 \eta_1 \left(1-q_{t}^{\gamma}\right)\right)\ ,
\end{split}
\end{equation}
 with 
\[
\mathbb{E}\left(d{\mathcal N}_{t}^{1,\gamma}\vert q_{t}^{\gamma}\right)=\gamma_1 \eta_1 \left(1-q_{t}^{\gamma}\right)dt, \quad 
d{\mathcal N}_{t}^{1,\gamma}d{\mathcal N}_{t}^{1,\gamma}=d{\mathcal N}_{t}^{1,\gamma} \ .
\]
  
In Fig.~\ref{fig:traj-thermal} we show  plots of the trajectory of the process for different values of $\gamma_1$.   

\subsection{Collapse-Measurement setup: Competition between two measurements }
We choose here $H=L^{th}=0$ and $p=2$ with two non-commuting reset measurement operators: the collapse measurement operator $N_1=\vert -\rangle \langle -\vert $, with rate $\gamma_1$,  and  the damping-spontaneous emission operator $N_2$ defined by
\begin{equation}
\label{eq:N2-007}
N_{2}=\left(\begin{array}{cc}
0 & 1\\
0 & 0
\end{array}  \right) = \vert + \rangle \langle - \vert,
\end{equation}
acting with rate $\gamma_2$.

Observe that $N_2$ does not commute with $N_1$, so QND property is broken. Note that not only is it  non-diagonal in the basis $\vert + \rangle, \vert - \rangle$, the matrix $N_2$ is in fact not diagonal in any basis. This means that even ignoring the $N_1$-measurement, the measurement of $N_2$ does not satisfy the QND hypothesis and, therefore, is not by itself an a priori collapse dynamics. This is why we use the generic word ``Measurement'' in the title of this section.

Anyway,  since $N_2$ is of rank one, the term $M_{N_2}[\rho_t^\gamma]\,  \dot {\mathcal N}_t^{2,\gamma}$ is a resetting dynamics to $\vert + \rangle \langle + \vert$. This is because
\begin{equation}
\label{eq:resetting2}
\rho \rightarrow \rho+M_{N_2}[\rho] = \cfrac{N_2 \rho N_2^{\dagger}}{{\rm{Tr}}\left(N_2 \rho N_2^{\dagger}\right)}=\vert +\rangle \langle +\vert \ .
\end{equation}
Hence we have here a competition between a resetting dynamics to $\vert - \rangle \langle - \vert$  induced by $N_1$ and a  resetting dynamics to $\vert + \rangle \langle + \vert$  induced by $N_2$. 
{Note that in Section \ref{sec:conjecture}, we will discuss extension to more general choice of  operator $N_2$.} 

\medskip

 As before we use the parameterisation given in Eq.~\eqref{eq:parameter1}. The SME Eq.~\eqref{eq:EMS-3} becomes then autonomous for the diagonal (see Eq.~\eqref{eq:sde-2-1-2-1-10} and Appendix \ref{sec:Appendix-:new})
\begin{equation}
\label{eq:sde-2-2-1-1}
\begin{split}
\dot q_{t}^{\gamma}& =-q_{t}^{\gamma}\left(\dot{\mathcal N}_{t}^{1,\gamma}-\gamma_{1}\eta_{1}\left(1-q_{t}^{\gamma}\right)\right) \\
&+\gamma_{2}\left(1-q_{t}^{\gamma}\right) +\left(1-q_{t}^{\gamma}\right)\left(\dot{\mathcal N}_{t}^{2,\gamma} -\gamma_{2}\eta_{2}\left(1-q_{t}^{\gamma}\right)\right),
\end{split}
\end{equation}
with
\begin{equation*}
\begin {split}
\mathbb{E}\left( d{\mathcal N}_{t}^{k,\gamma}\vert q_{t}^{\gamma}\right)=\gamma_{k} \eta_k \left(1-q_{t}^{\gamma}\right)dt \ , \\
 d{\mathcal N}_{t}^{k,\gamma} d{\mathcal N}_{t}^{\ell,\gamma}=\delta_{k\ell} \ d{\mathcal N}_{t}^{k,\gamma} \ .
\end{split}
\end{equation*}

In Fig.~\ref{fig:traj-coll}, we show plots of the trajectory of the Collapse-Measurement setup for different values of $\gamma_1$.

\section{Summary of main results in the Precise setup}
\label{sec:mainresults}

\begin{figure*}
    \centering
    \includegraphics[scale=0.5]{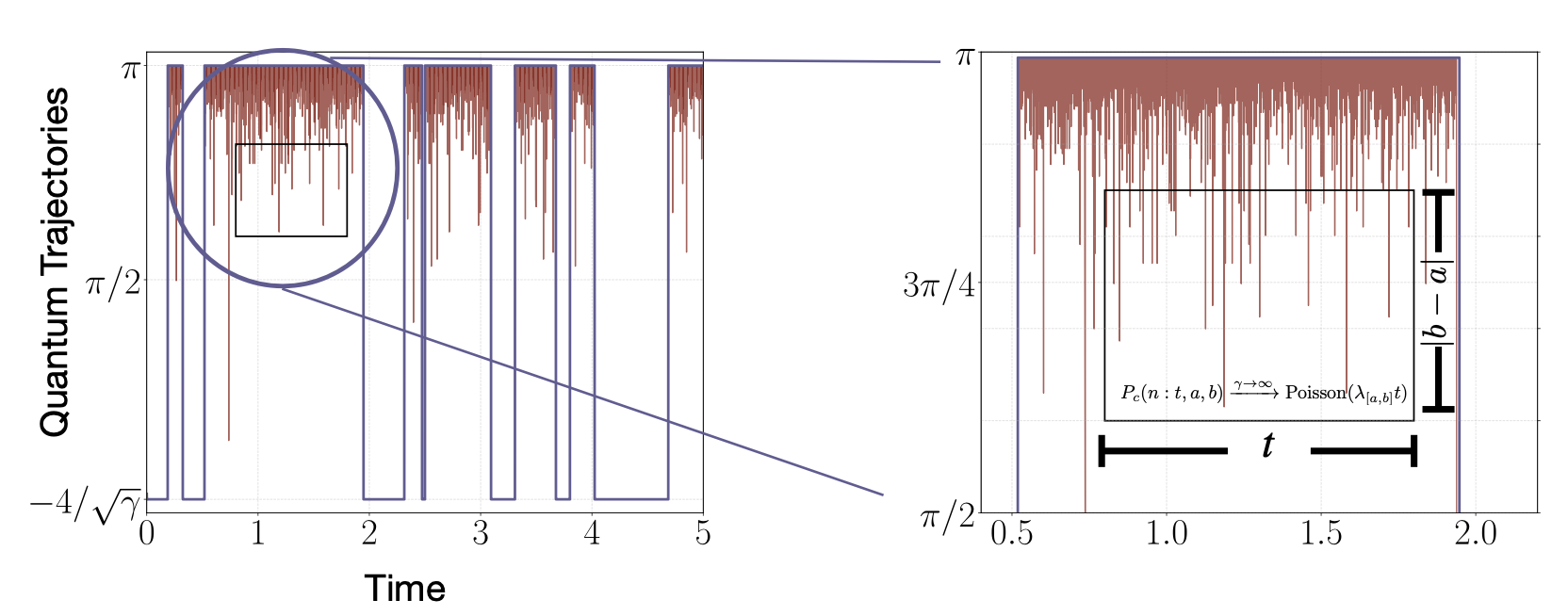}
\caption{Pictorial depiction of the construction of   $P_c(n:t,a,b)$: the probability of observing \( n \) pre-spikes, given that there are no jumps in between. Quantum trajectories are characterized by quantum jumps (shown in violet) and are accompanied by spikes (shown in brown). The spike statistics occurring between consecutive jumps within a black box of width \( |b-a| \) and length \( t \) is the focus of this article. It is anticipated that at large measurement strength, these spike statistics follows a Poisson distribution.}
\label{fig:schematic}
\end{figure*}

From numerical simulations (see Figs. \ref{fig:traj-rabi}, \ref{fig:traj-thermal} and \ref{fig:traj-coll}),  we observe that, in the large $\gamma$ limit, a typical trajectory is composed of two parts. 

The first part is given by a pure jump Markov process with state space{\footnote{The fact that the state space is $\{0,\pi\}$ and not $\{0,1\}$ is due to the change of variable done in Eq.~\eqref{eq:QIS-1}. }} $\{0,\pi\}$ (in the Collapse-Unitary setup) or $\{0,1\}$ (in the Collapse-Thermal and Collapse-Measurement setups), corresponding to quantum jumps.

The second part is a decoration of this trajectory by a large number of one-sided vertical lines connecting $\pi$ in the Collapse-Unitary setup (resp. $0$ in the Collapse-Thermal and Collapse-Measurement setups) to random values  in $(0,\pi)$ (resp.  $(0,1)$ in the Collapse-Thermal and Collapse-Measurement setups) at random times. These vertical lines correspond in fact to the deterministic trajectories (we call them pre-spikes) during successive clicks in the strong measurement limit. Our main goal now is to describe the statistics of the clicks in the strong measurement regime.  {Since the spikes originate only on the part of the quantum jump trajectory equal to $\pi$ for the Collapse-Unitary case and $0$ for the other two cases, we will focus on respectively these precise initial conditions. }

As seen in Figs.~ \ref{fig:traj-rabi}, \ref{fig:traj-thermal},\ref{fig:traj-coll}, for any finite $\gamma$, the pre-spikes have a specific structure consisting of a deterministic flow and a sudden reset to a particular state. In the special limit $\gamma \to \infty$, the pre-spikes become infinitely thin and occur with an infinite density (in time), and have a distribution of heights. Hence, we have a clear picture of how spikes emerge from pre-spikes. To characterize the statistic of spikes precisely, we define a spike to be localized at time $t$, if its tip ends at a spatial point $x$. We then ask for the number of spikes observed in a two-dimensional space-time box of width $(0,t)$ in time and $(a,b)$ in space.  

The probability of observing $n$ pre-spikes, conditioned on no jumps (i.e. no pre-spikes whose tips extend to the opposite end, see Fig.~\ref{fig:schematic}) occurring during this time window, will be denoted by $P_c(n:t,a,b)$. To be more precise, as explained above, we focus only on initial conditions for which we can observe pre-spikes before a quantum jump,  i.e. $\pi$ in the Collapse-Unitary setup and $0$ in the other cases. Hence, we describe the non-trivial spike statistics only during the (random) period of time in which the investigated process starts from a pointer state with spikes and does not jump to a new pointer state (without spikes) during this time period. By Markov property our analysis is in fact valid for any period of time where we really have spikes.

{One of our main result of this paper is the proof, in the limit of large $\gamma$,  of the Poisson distribution: }

\begin{equation}
\label{eq:spikes-stat}
\begin{split}
&\lim_{\gamma\rightarrow\infty}P_c(n:t,a,b)=\tfrac{\left(\lambda_{\left[a,b\right]}t\right)^{n}}{n!}\; \exp\left(-\lambda_{\left[a,b\right]}t\right) \\
& \quad \quad {\textrm{where}}~~ \lambda_{\left[a,b\right]}=\int_{a}^{b}dx \, I(x) \ ,
\end{split}
\end{equation}
and the intensity for the three cases are respectively given by:
\begin{align}
I(x)&=   \frac{4 \omega \sin(x/2)}{\cos^3 (x/2)} ~~~~{\rm Collapse-Unitary}\nonumber \\    
&=   \frac{W_{-,+}}{x^2}   ~~~~~~~~~~~{\rm Collapse-Thermal}\nonumber  \\    
 &=   \frac{\gamma_2(1-\eta_2)}{x^2} ~~~~~~{\rm Collapse-Measurement} \ .
\label{eq:spikeintense}
 \end{align}

\begin{figure*}
 	\centering
 \includegraphics[width=0.43\linewidth,height=0.29\linewidth]{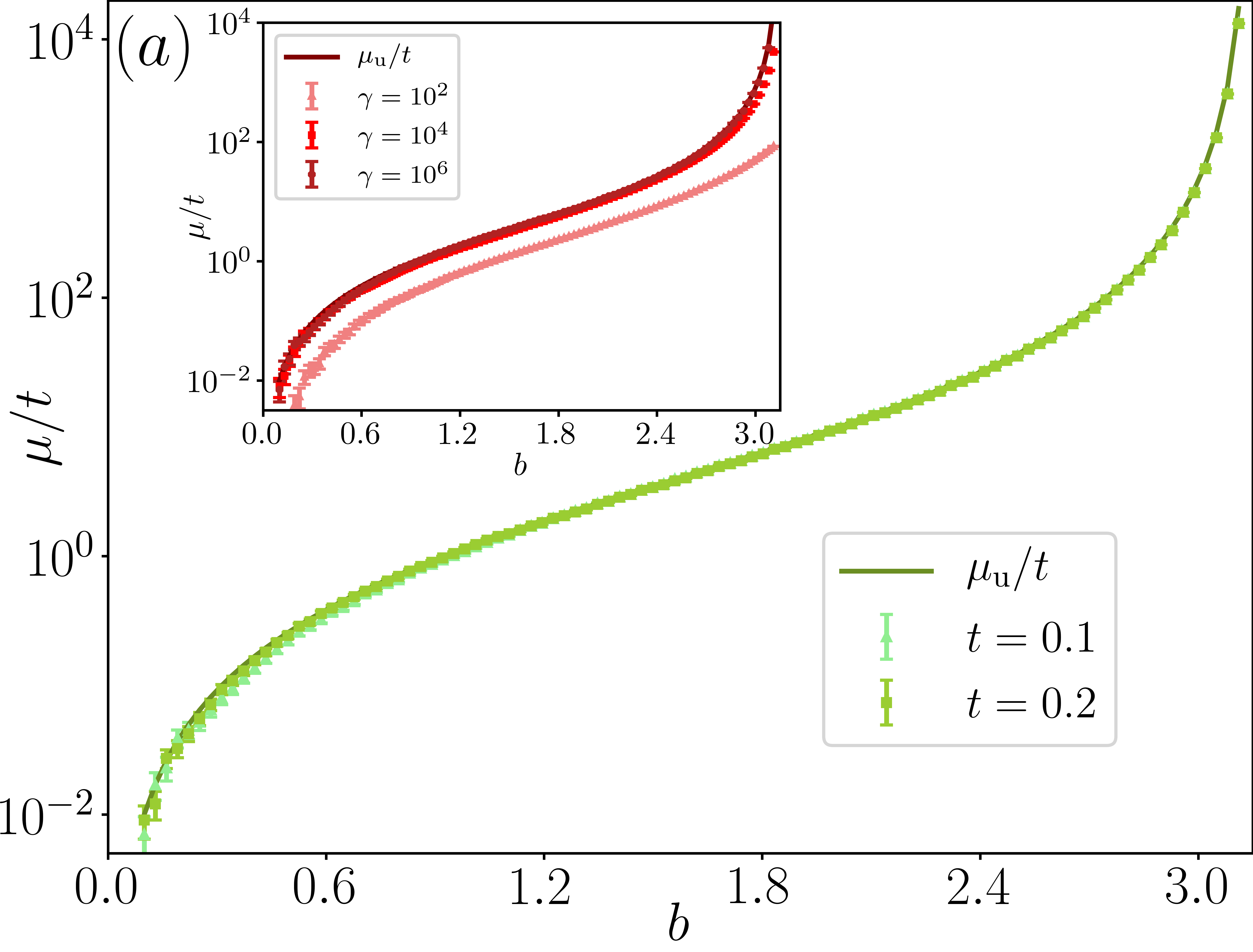}\hspace{0.8cm}
\includegraphics[width=0.43\linewidth,height=0.29\linewidth]{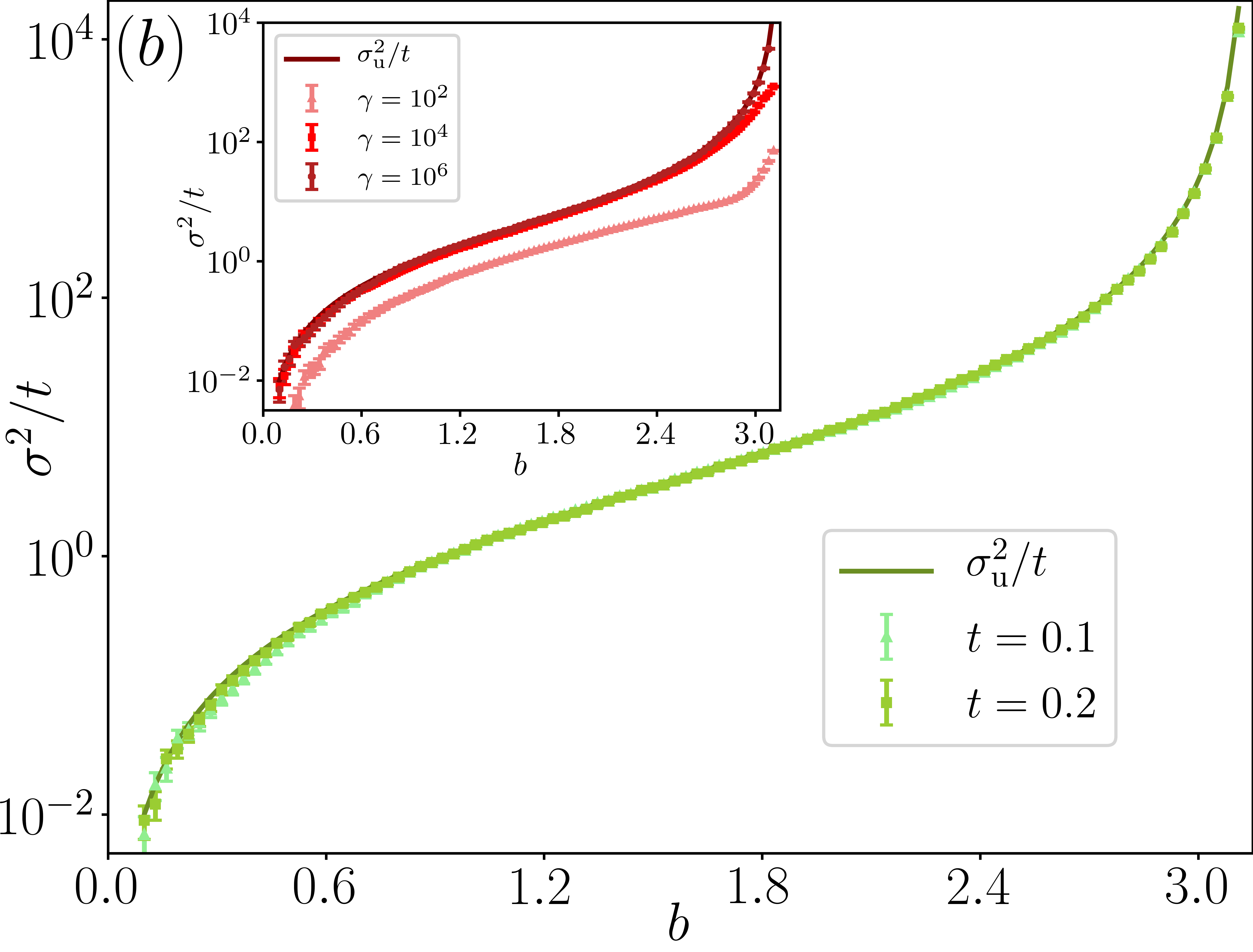}
 	\caption{Collapse-Unitary case: (a) Mean (b) Variance of the distribution (conditioned on no-jumps) of the number of spikes (from $\theta_t=\pi$) per unit time in a space-time box $(0,b)\times(0,t)$ plotted against the box edge $b$ for $\gamma=10^6$. The data-points were obtained for $\omega=1$ by averaging over $10^4$ realizations. $\mu_\mathrm{u}$ and $\sigma^2_\mathrm{u}$ are the mean and variance, respectively, of the theoretically-predicted Poisson process in Eq.~\eqref{eq:spikes-stat}. The data in the insets, plotted for a fixed time $t=0.1$, converge to $\mu_\mathrm{u}/t$ and $\sigma^2_\mathrm{u}/t$ at large $\gamma$.}
 	\label{plts:Rabi}
 \end{figure*}

\begin{figure*}
 	\centering
 \includegraphics[width=0.43\linewidth,height=0.29\linewidth]{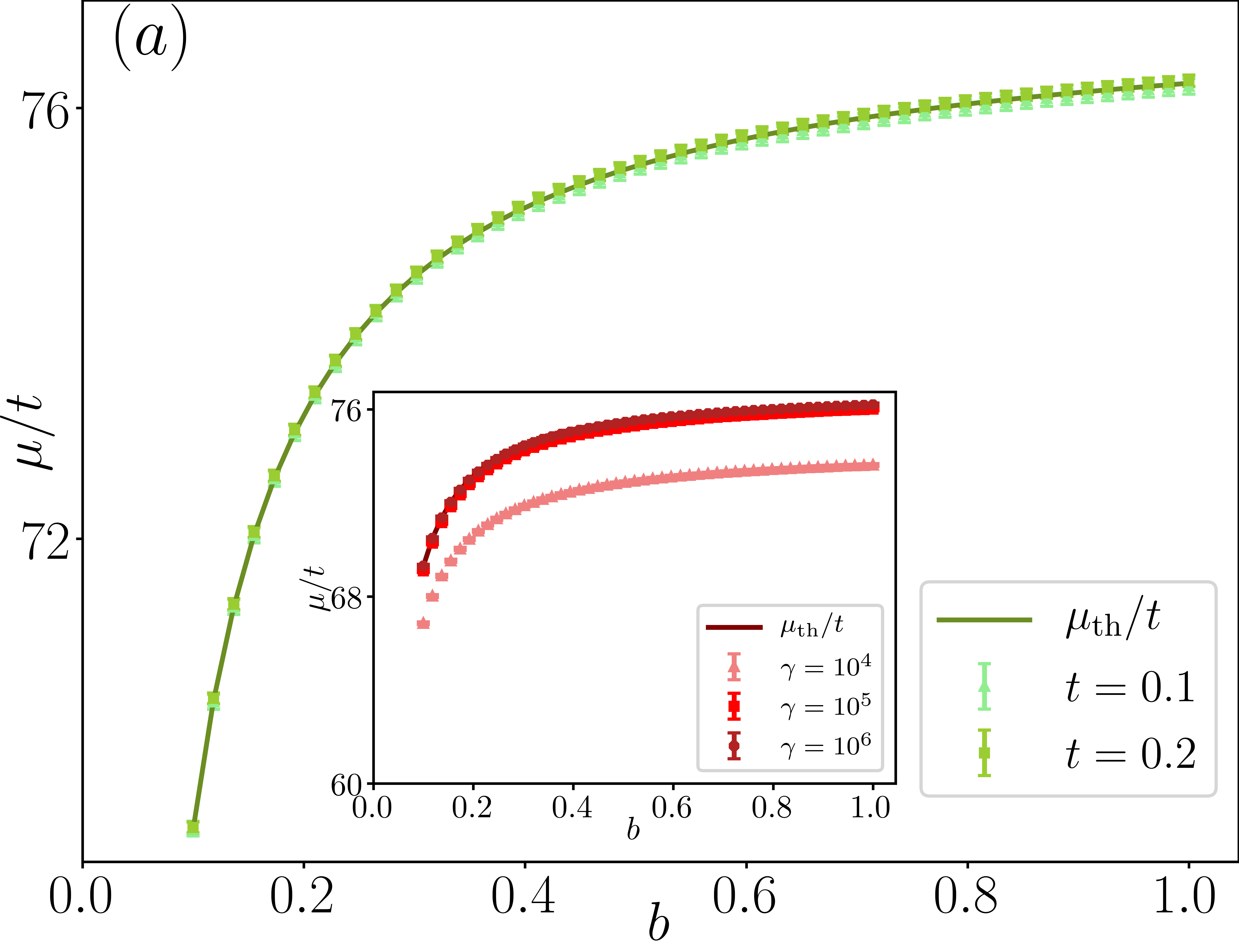}\hspace{0.8cm}
\includegraphics[width=0.43\linewidth,height=0.29\linewidth]{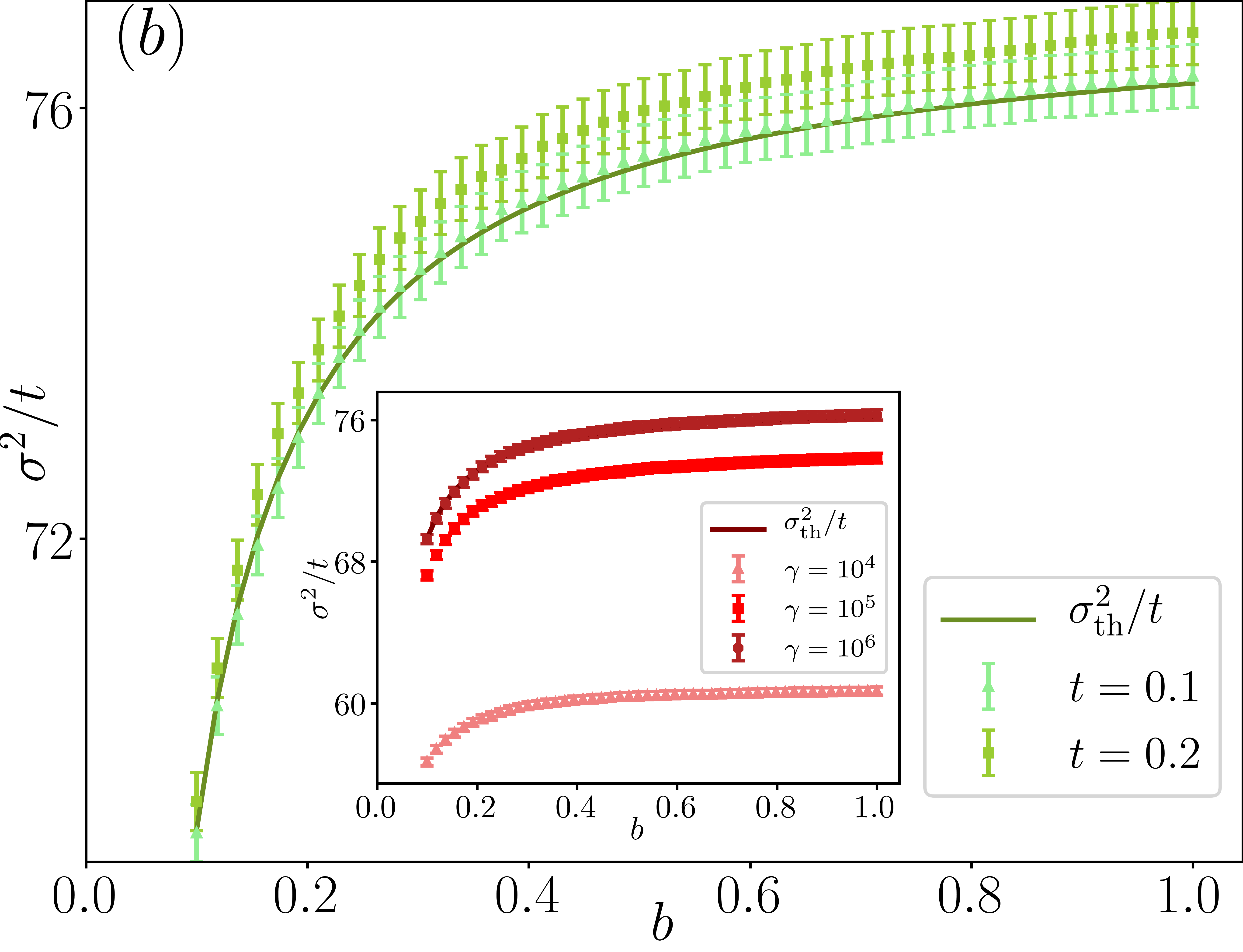}
 	\caption{Collapse-Thermal case: (a) Mean (b) Variance of the distribution (conditioned on no-jumps) of the number of spikes (from $q_t=0$) per unit time in a space-time box $(0.01,b)\times(0,t)$ plotted against the box edge $b$ for $\gamma=10^6$. The data-points were obtained for the parameters $W_{-,+}=0.77$, $W_{+,-}=0.23$ by averaging over $10^5$ realizations. $\mu_\mathrm{th}$ and $\sigma^2_\mathrm{th}$ are the mean and variance, respectively, of the theoretically-predicted Poisson process in Eq.~\eqref{eq:spikes-stat}. The data in the insets, plotted for a fixed time $t=0.1$, converge to $\mu_\mathrm{th}/t$ and $\sigma^2_\mathrm{th}/t$ at large $\gamma$.}
 	\label{plts:Thermal}
 \end{figure*}

\begin{figure*}
 	\centering
 \includegraphics[width=0.43\linewidth,height=0.29\linewidth]{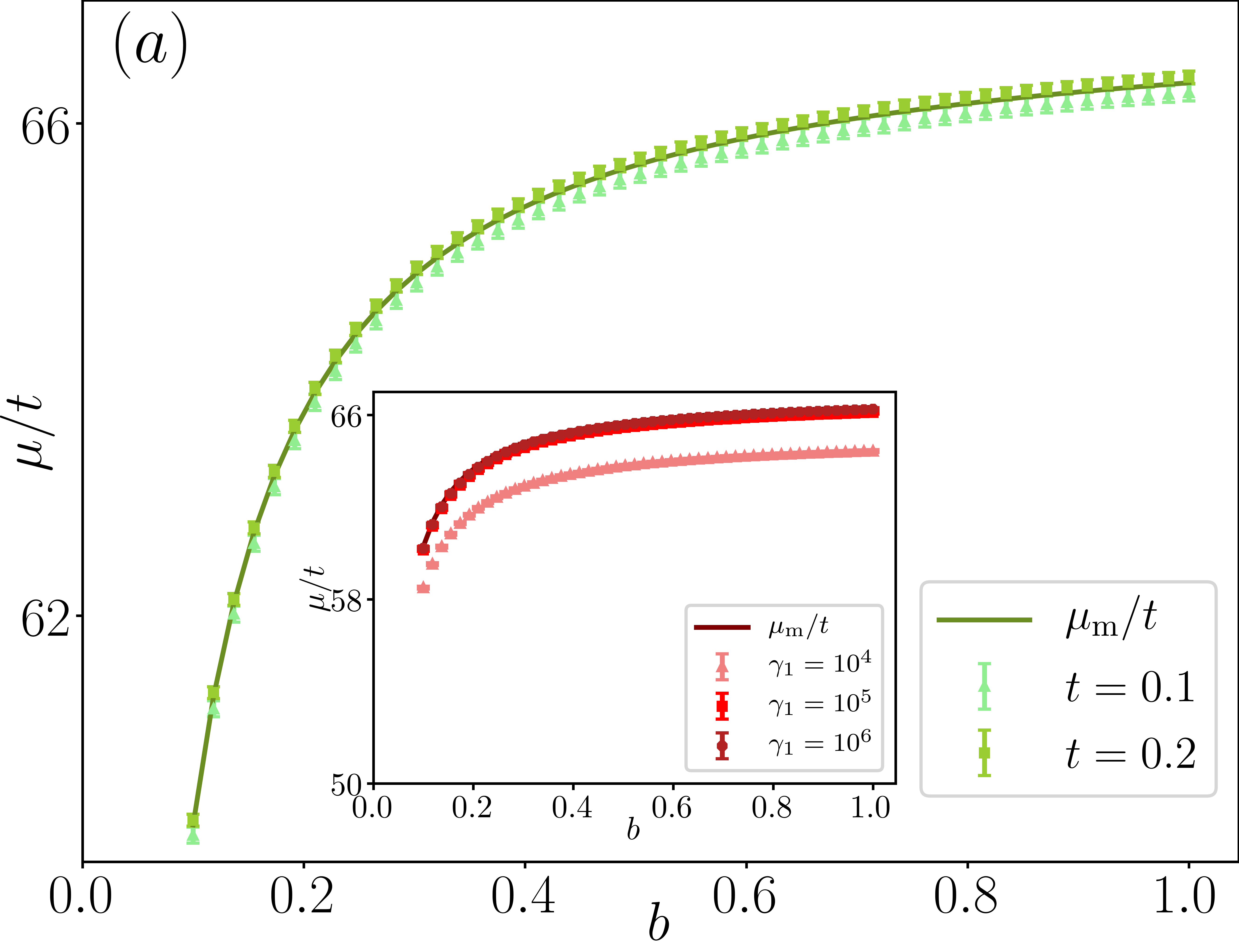}\hspace{0.8cm}
\includegraphics[width=0.43\linewidth,height=0.29\linewidth]{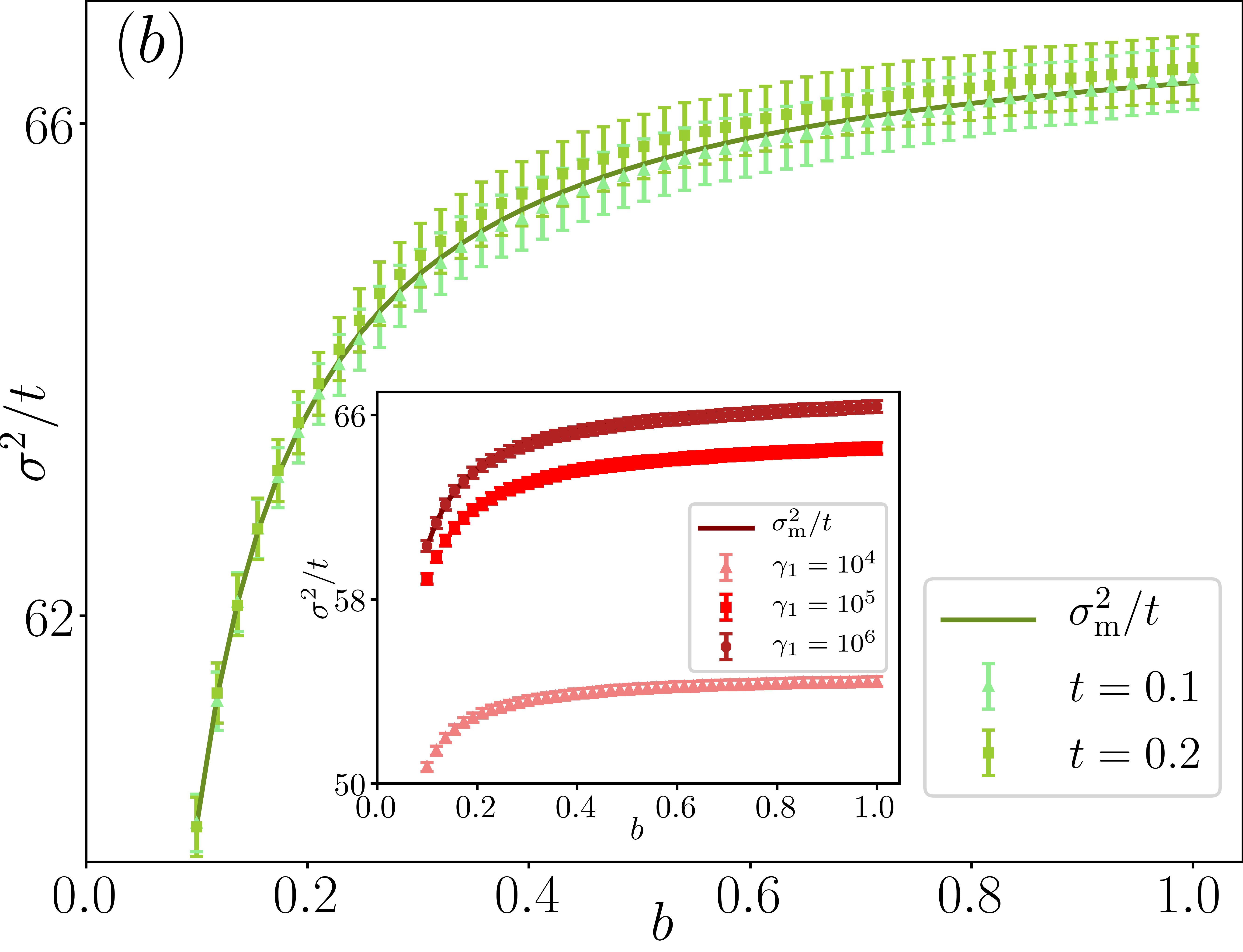}
 	\caption{Collapse-Measurement case: (a) Mean (b) Variance of the distribution (conditioned on no-jumps) of the number of spikes (from $q_t=0$) per unit time in a space-time box $(0.01,b)\times(0,t)$ plotted against the box edge $b$ for $\gamma_1=10^6$. The data-points were obtained for the parameters $\gamma_2=1$ and $\eta_1=\eta_2=0.33$ by averaging over $10^5$ realizations. $\mu_\mathrm{m}$ and $\sigma^2_\mathrm{m}$ are the mean and variance, respectively, of the theoretically-predicted Poisson process in Eq.~\eqref{eq:spikes-stat}. The data in the insets, plotted for a fixed time $t=0.1$, converge to $\mu_\mathrm{m}/t$ and $\sigma^2_\mathrm{m}/t$ at large $\gamma_1$.}
 	\label{plts:Collapse}
 \end{figure*}

\begin{figure}
    \centering
    \includegraphics[width=0.9\linewidth]{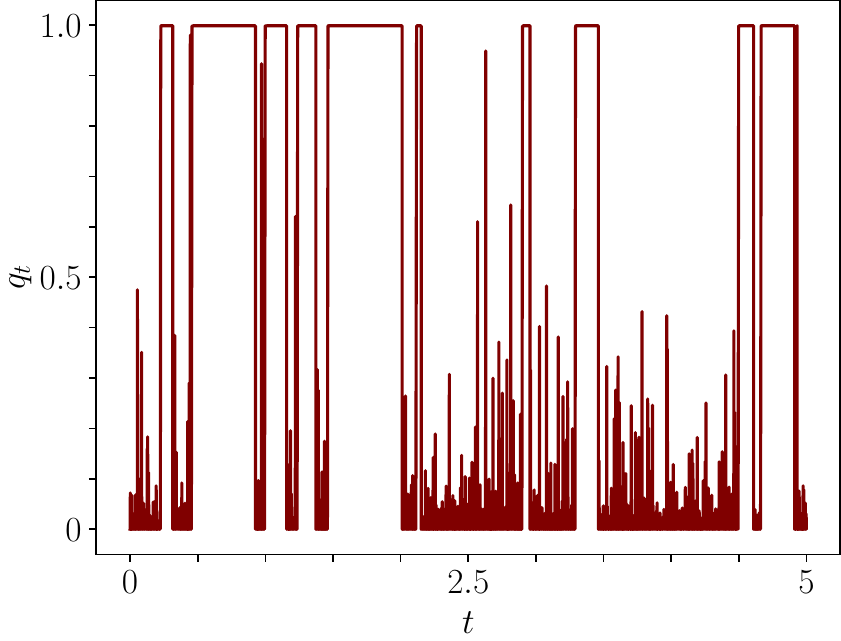}
    \caption{Typical trajectories for Collapse-Unitary setup for inefficient measurements, for the parameters $\eta_1=0.33$, $\gamma=10^4$ and $\omega=1$. Similar to the Collapse-Thermal and Collapse-Measurement cases, the jumps between $q_t=0$ and $q_t=1$ are interspersed with spikes from $q_t=0$ in the strong measurement limit.}
    \label{fig:RabiInefficient}
\end{figure}

Some comments are the following:
\begin{itemize}
\item In Figs.~\ref{plts:Rabi},\ref{plts:Thermal},\ref{plts:Collapse} we show a comparison of the above analytic forms of the three processes  with  computations of the mean and variance of the spike statistics obtained from numerical simulations. The Poisson property is illustrated by the overlap of the empirical mean and the empirical variance. In all cases we see excellent agreement of the numerics with the analytic predictions.

\item {The first result, in the Collapse-Unitary setup,  is proved only for the fully efficient case ($\eta_1=1$) while the latter two are valid for any non-vanishing efficiency ($0<\eta_1\leq 1$). 
Anyway, numerical simulation of the inefficient case ($0<\eta_1< 1$) in the Collapse-Unitary set-up (i.e.  equation Eq.~\eqref{eq:sde-2}) show that the spikes exist anyway in the non-autonomous $q$ process (See Fig. \ref{fig:RabiInefficient}).}

\item  By a change of variable $q=(1+\cos x)/2$ for the first case, all three intensities acquire the same universal form. This change of variable is in fact natural since it corresponds to returning to the representation of the qubit in Eq.~\eqref{eq:parameter1} with the inverse of the change of variable given by Eq.~\eqref{eq:QIS-1}.
 Interestingly, the spikes statistics above are similar to the spikes statistics obtained in \cite{Bauer,Tilloy,Bauer2018,BCCNP2023}, in the case where the noise $\mathcal N$ is not Poissonian, but Gaussian. 
 
\item {In the Collapse-Measurment setup,  for a perfect measurement efficiency of $N_2$ (i.e. $\eta_2=1$), the spikes disappear completely, indicating a complex interplay of measurement of two non-commuting observables in non-QND measurement.}

\item {In these three cases, the jump transition rates of the jump process} can be computed explicitly. In the Collapse-Unitary setup, the jumps between the two pointer states occur with rate $4 \omega$ while in the Collapse-Thermal setup, the jumps between the states occur with rates $W_{-,+}$ and $W_{+,-}$ respectively, and in the Collapse-Measurement setup, only transitions are one-sided with rate $\gamma_2$ (parameters are specified in Sec. \ref{sec:precisesetup}).
\end{itemize} 

\section{Mathematical Derivations}
\label{sec:derivations}
\subsection{Study in the Collapse-Unitary setup}
\label{sec:an-CUsetup}

To simplify notation, we denote $\gamma_1$ by $\gamma$ and ${\tilde{\mathcal  N}}^{1, \gamma}$ by ${\mathcal N}$, $\theta^\gamma$ by $\theta$. We also choose $k_\gamma =\sqrt {\omega\gamma}$ in order to get quantum spikes in the large $\gamma$ limit (see Appendix \ref{app:Rabi-scaling} for justification). The dynamics is then specified by the following stochastic evolution equation (see Eq.~\eqref{eq:sde-angle-1-1-1-1}):
\begin{equation}
\label{eq:SDEMain}
d\theta_{t}=\Omega \left(\theta_{t}\right)dt+\left(\pi-\theta_{t}\right)d{\mathcal N}_{t} \ ,
\end{equation}
where the drift term $\Omega:=\Omega^\gamma$ and the Poissonian noise $\mathcal N$ satisfy:
\begin{align}
\Omega (\theta) &=-2\sqrt{\omega\gamma} \left[1+ \sqrt{\tfrac{\gamma}{16\omega}} \sin\theta\right], \\
\mathbb{E}\left[d{\mathcal N}_{t} \vert \theta_t  \right] &=\alpha(\theta_t)\,dt:=\gamma\sin^2\tfrac{\theta_t}{2}\,dt \ . 
\end{align} 
Hence the dynamics evolves deterministically in $(\theta^*=-\sin^{-1}(4\sqrt{\omega/\gamma}),\pi)$ but at some random times prescribed by $\mathcal N$, the angle is resetting to the value $\pi$, i.e., a ''click'' occurs. Rarely, during the deterministic evolution, no resetting occurs and the trajectory is able to reach $\theta^*$ and is stuck around $\theta^*$ for a random time until a new resetting to $\pi$ occurs. In the large $\gamma$ limit, as $\theta^* \to 0$,  we say then that we have a jump from $\pi$ to $0$ and afterwards a jump (instantaneous) from $0$ to $\pi$.

With the notations of \cite{DCD2023}, the situation considered here corresponds then to the special case $\gamma_0=\sqrt{\omega\gamma}$, or equivalently $\lambda=\gamma/(4 \gamma_0)=\sqrt{\tfrac{\gamma}{16\omega}}$. We recall that we are interested in the strong measurement regime $\gamma \to \infty$, not investigated in \cite{DCD2023}. Hence, the angle ${\hat \theta}_t$, evolving according to the deterministic equation 
\begin{equation}
\label{eq:detflow}
d {\hat \theta}_t = \Omega ({\hat \theta}_t) \, dt \ ,
\end{equation}
is confined to the range $(-\sin^{-1}(1/\lambda),\pi) \approx (-4\sqrt{\omega/\gamma},\pi)$,
the last approximation being valid since $\gamma$ is large. In particular, for $t \ge 0$, we have the exact solution \cite{DCD2023}:
\begin{align}
\label{detsol1}
\tan \left(\frac{{\hat \theta}_{t}(0 \vert \pi)}{2} \right)=-\frac{\sinh(\beta\sqrt{\omega\gamma}\ t-\phi)}{\sinh(\beta \sqrt{\omega\gamma} \ t)} \ ,
\end{align}
where 
\begin{equation}
\label{eq:beta}
    \beta= \sqrt{\frac{\gamma}{16\omega}-1} =\sinh(\phi) \ .
\end{equation}

Let $\tau$ be the time taken by a deterministic trajectory $\hat\theta$ to evolve from ${\hat \theta}_0=\pi$ to ${\hat \theta}_\tau=0$. From Eq.~\eqref{detsol1}, we note $\tau=\phi/(\beta\sqrt{\omega\gamma})$. For $\gamma\gg1$,  
    $\tau\approx2\log{\gamma}/\gamma$.
If $c\in (0,\pi)$, let us also denote the time for the deterministic dynamics to reach $c$ when starting from $\pi$ by $\tau_c$. A straightforward inversion of Eq.~\ref{detsol1} yields 
\begin{equation}
\label{eq:tau_c}
    \tau_c = \frac{1}{2\beta\sqrt{\omega\gamma}}\log\left[1+\frac{2\beta}{\tan(c/2)+e^{-\phi}}\right] \ .  
\end{equation}

Given that the initial state is $\theta_0$, we consider the probability $p_0^t(t_1,t_2,\ldots,t_n|\theta_0)$ of observing no jumps and a sequence of $n$ clicks occurring exactly at times $t_1 < t_2< \ldots t_n$ within the time interval $(0,t)$. The no-jump condition is enforced by the inequality $\tau\ge t_j -t_{j-1}$ for any $j \le n$. The quantum trajectory is deterministic between the clicks and the sequence $\{\theta_0,\theta_{t_1},\pi,\theta_{t_2},\pi,\ldots,\theta_{t_n},\pi,\theta_t\}$ determines the quantum trajectory. The evolution from $\theta_0\to \theta_{t_1}$, $\pi \to \theta_{t_i}$ (for $i=2,\ldots,n$) and $\pi \to \theta_t$ is given by the deterministic flow in Eq.~\eqref{eq:detflow}.

Let us define $\mu(\Delta t|\theta_i)$ as the probability to not have a click during the time interval $\Delta t$ for the initial condition $\theta=\theta_i$. This is given by
\begin{equation}
\label{eq:noclick}
\begin{split}
    \mu(\Delta t|\theta_i) & = e^{-\int_0^{\Delta t} ds \alpha({\hat \theta}_s(0|\theta_i))} =\tfrac{\Omega(\theta_{i})}{\Omega({\hat \theta}_{\Delta t}(0 \vert\theta_{i}))}\,{\rm e}^{-\tfrac{\gamma \Delta t}{2}} \  .
\end{split}    
\end{equation}
Then we have:
\begin{equation}
\begin{split}
&p_0^t(t_1,t_2,\ldots,t_n|\hat\theta_0)\\
&= \mu(t_1|\theta_0) \alpha (\hat\theta_{t_1}) ~\prod_{i=2}^n [\mu(t_i-t_{i-1}|\pi) \alpha (\hat\theta_{t_i-t_{i-1}}(0|\pi)] \\
&\times \mu(t-t_n|\pi) \  .
\label{eq:p0}
\end{split} 
\end{equation}
Using the explicit solution in Eq.~\eqref{detsol1} and  after some algebra we get:
\begin{equation}
\label{eq:mualpha-unit}
\begin{split}
    &\mu( t |\pi) ~\alpha(\hat\theta_{t}(0|\pi)) = \frac{\gamma \sinh^2 (\beta \sqrt{\omega\gamma} t - \phi)}{\beta^2} \ ,\\
    &\mu(t |\pi)  = \frac{\sinh^2 (\beta \sqrt{\omega\gamma} t) + \sinh^2 (\beta \sqrt{\omega\gamma} t - \phi)}{\beta^2} \ . 
    \end{split}
\end{equation}

The joint probability $P_j(n:t,a,b)$ to observe exactly $n$ pre-spikes in the time interval $[0,t]$ and in the space interval $[a,b]$ (where $0<a<b<\pi$), starting from $\pi$, and such that no jumps occur in the time interval $[0,t]$, is given by 
\begin{equation}
\begin{split}
&P_j(n:t,a,b)\\
&= \sum_{m=0}^\infty \frac{(n+m)!}{n! m!} \prod_{i=1}^{n+m} \int_0^{t_{i+1}} dt_i  ~ p_0^t(t_1,t_2,\ldots,t_{n+m}|\pi) \\
&\times \prod_{j=1}^m [~\Theta(\tau_b- (t_{j}-t_{j-1}))  + \Theta((t_{j}-t_{j-1})-\tau_a)~]\\ 
&\times \prod_{j=1}^n  \Theta((t_{j}-t_{j-1})-\tau_b) ~ \Theta(\tau_a-(t_{j}-t_{j-1}))\\
&\times \prod_{j=1}^{n+m} \Theta(\tau- (t_{j}-t_{j-1}))~\Theta(\tau-(t-t_{n+m})) \ , 
\end{split}
\label{eq:Pc}
\end{equation}
where we have defined $t_{n+m+1}=t$ and $\Theta$ is the Heaviside function.
Taking time-Laplace transform $\hat{P}_j(n:\sigma,a,b)= \int_{0}^{\infty} dt e^{-\sigma t} P_j(n:t,a,b)$ gives
\begin{equation}
\label{eq:Laplce-cond}
   \hat{P}_j(n:\sigma,a,b)= \sum_{m=0}^\infty \frac{(n+m)!}{n! m!} 
   C^m(\sigma) D^n(\sigma) E(\sigma) \ ,
\end{equation}
where 
\begin{equation}
\begin{split}
C(\sigma)&=\int_0^\infty dt ~\mu(t|\pi) ~\alpha(\hat\theta_t(0 \vert \pi))~ [ \Theta(\tau_b-t) + \Theta (t-\tau_a)]\\
& \quad \quad \quad \quad \times ~ \Theta (\tau-t) ~e^{-\sigma t}\ , \\ 
D(\sigma)&=\int_0^\infty dt ~\mu(t|\pi) ~\alpha(\hat\theta_t(0 \vert\pi))~\Theta(\tau_a-t) \Theta (t-\tau_b)  e^{-\sigma t} \ , \\
E(\sigma)&=\int_0^\infty dt~ \mu(t|\pi)~ \Theta(\tau-t)~ e^{-\sigma t} \ .
\label{eq:CDE}
\end{split}
\end{equation}
Performing the summation, we get
\begin{align}
   \hat{P}_j(n:\sigma,a,b)= \frac{  D^n(\sigma) E(\sigma) }{[1-C(\sigma)]^{1+n}} \ . 
   \end{align}
Writing the generating function ($0\le s \le 1$)
\begin{equation}
Z(s:\sigma,a,b)=\sum_{n=0}^\infty s^n \hat{P}_j(n:\sigma,a,b) \ ,
\label{eq:Zc}
\end{equation}
we find the exact formula
\begin{align}
   \label{eq:functionZ}
   Z(s:\sigma,a,b)= \frac{ E(\sigma) }{1-C(\sigma)-sD (\sigma) } \ . 
   \end{align}

Thanks to the explicit expressions Eq.~\eqref{eq:mualpha-unit}, the functions $C(\sigma)$, $D(\sigma)$ and $E(\sigma)$ can be computed explicitly and their asymptotic behaviour for large $\gamma$ can be obtained (See Appendix \ref{app:LaplaceTransform} for details). It follows that for any $s\in [0,1]$ fixed we have in the large $\gamma$ limit that
\begin{equation}
\label{eq:Z_Rabi}
\begin{split}
     Z&(s:\sigma,a,b)\\
     &\approx \cfrac{1}{\sigma+4\omega+4\omega(1-s) [\tan^2(b/2) -\tan^2 (a/2)]} \ .
\end{split}
\end{equation}

The probability $S(t)$ of no jump occurring from $\pi$ to $0$ in the time interval $(0,t)$ is obtained by taking $s=1$ in the generating function in Eq.~\eqref{eq:Z_Rabi}, which yields $S(t)=e^{-4\omega t}$. Since $1-S(t)$ is the cumulative distribution function of the distribution of time of first passage from $\pi$ to $0$, this implies that the first time to reach $0$ is exponentially distributed with a mean equal to $1/(4\omega)$. The probability $P_c(n : t, a, b)$ to observe exactly $n$
pre-spikes in the time interval $[0, t]\times[a, b]$, \textit{given} that no
jump occurs in the time interval $[0, t]$, can be obtained by 
\begin{equation*}
P_c(n:t,a,b)=\frac{P_j(n:t,a,b)}{S(t)}\,,    
\end{equation*} 
which is given by the Poisson distribution in the large $\gamma$ limit (as announced in Eq.~\eqref{eq:spikes-stat}),
\begin{equation}
        \frac{(\lambda_{[a,b]} t)^ne^{-\lambda_{[a,b]}t}}{n!},
  \hspace{0.2cm}\text{where}\hspace{0.1cm}\lambda_{[a,b]} =\int_a^bdx\,\tfrac{4\omega\sin(x/2)}{\cos^3 (x/2)} \ .    
\end{equation}

Moreover, by using Eq.~\eqref{eq:noclick} and \eqref{detsol1} we have that the probability to have no resetting to $\pi$ by starting from $0$ in a time interval of length $\Delta t$ is given by 
\begin{equation}
    \mu (\Delta t \vert 0) = \ \cfrac{-2 \sqrt{\omega\gamma} e^{-\tfrac{\gamma \Delta t}{2}}}{\Omega \left( - 2 \tan^{-1} \left( \tfrac{\sinh (\beta \sqrt{\omega\gamma} \Delta t)}{\sinh (\beta \sqrt{ \omega\gamma} \Delta t + \phi) }\right) \right)} \ .
\end{equation}
By performing the expansion in $\gamma$ we get that 
\begin{equation}
    \mu (\Delta t \vert 0) \approx \exp\left( - 4\omega\Delta t \right),
\end{equation}
so that the time to be resetting to $\pi$  when starting from $0$ is also exponentially distributed with mean $1/(4\omega)$.

\bigskip

\subsection{Study in the Collapse-Thermal setup}

To simplify notation, we denote $\gamma_1$ by $\gamma$ and ${\mathcal N}^{1, \gamma}$ by ${\mathcal N}$, $q_t^\gamma$ by $q_t$. Eq.~\eqref{eq:EMS-3-1-2-1-1-1-2} can then be rewritten as
\begin{equation}
\label{eq:SDEMain-th}
dq_{t}=\Omega\left(q_{t}\right)dt-q_{t}\, d{\mathcal N}_{t} \ ,
\end{equation}
where the quadratic drift term $\Omega$ and the rate function $\alpha$ of ${\mathcal N}_t$ are given by:
\begin{equation}
\begin{split}
&\Omega(q) =W_{-,+}-q(W_{+,-}+W_{-,+}-\gamma\eta)-\gamma\eta q^2 \ , \\
&\mathbb{E}\left[d{\mathcal N}_{t} \vert q_t \right] =\alpha(q_t)\,dt=\gamma\eta(1-q_t)\,dt \ .
\end{split} 
\end{equation}
The roots of $\Omega(q)$ are the fixed points of the no-click deterministic dynamics, given by 
\begin{equation*}
q_{\pm}=\tfrac{\gamma\eta-W_{+,-}-W_{-,+}\pm\sqrt{(W_{+,-}+W_{-,+}-\gamma\eta)^2+4W_{-,+}\gamma\eta}}{2\gamma\eta}\ ,
\end{equation*}
which are always real and distinct. Note that $q_-<0<q_+$, provided all the parameters are not identically zero. For an initial condition $q_0<q_+$, the stochastic dynamics is confined to $[0,q_+)$ for $\gamma\gg 1$.  
\\\\
The solution of the deterministic equation $d \hat{q}_t=\Omega (\hat q_t) dt $ starting from $q_0$ is given by
    \begin{align}
{\hat q}_t(q_0)&=\frac{q_+(q_0-q_-)+q_-(q_+-q_0)e^{-\gamma\eta(q_+-q_-)t}}{(q_0-q_-)+(q_+-q_0)e^{-\gamma\eta(q_+-q_-)t}} \ .
\label{eq:thmsol}
\end{align}

For $c \in [0,1]$, let $\tau_c$ be the time taken to evolve from $q_0=0$ to $q=c$. A straightforward inversion of Eq.~\eqref{eq:thmsol} shows that
\begin{equation}
 \label{tau_c-thm}
 \tau_c= \frac{1}{\eta\gamma(q_+-q_-)}\log\Bigg|\frac{(c-q_-)q_+}{q_-(q_+-c)}\Bigg| \ , 
 \end{equation}
and let $\tau$ be the time taken by a deterministic trajectory to evolve from $q=0$ to $q=1-\epsilon$. For $\gamma\gg1$, 
\begin{equation*}
    \tau\sim\frac{1}{\gamma\eta}\log\Bigg|\frac{\gamma^2\eta^2\left(1-\epsilon\right)}{W_{+,-}W_{-,+}+\epsilon\left(\gamma^2\eta^2-W_{+,-}\gamma\eta\right)}\Bigg| \ .
\end{equation*}

Similar to the Collapse-Unitary case, we explicitly obtain the functions

\begin{equation}
\label{eq:Alan007}
\begin{split}
& \mu(t|q_0) = e^{-\int_0^t ds \alpha[{\hat q}_s(q_0)]}\\
         &=\frac{(q_0-q_-)e^{-\gamma\eta(1-q_+)t}+(q_+-q_0)e^{-\gamma\eta(1-q_-)t}}{q_+-q_-}\ ,  \\
          &\mu(t|q_0) \alpha [q_t(q_0)]\\
    & =\tfrac{\gamma\eta}{q_+-q_-}\Big[(q_0-q_-)(1-q_+)e^{-\gamma\eta(1-q_+)t}+\\
    &\hspace{1.6cm}(q_+-q_0)(1-q_-)e^{-\gamma\eta(1-q_-)t}\Big] \ .
\end{split}    
\end{equation}
which we plug into the Eq.~\eqref{eq:CDE} to explicitly obtain the functions $C(\sigma)$, $D(\sigma)$ and $E(\sigma)$ and obtain their asymptotic behaviour for large $\gamma$ (See Appendix \ref{app:LaplaceTransform} for details).

\medskip
Consider the probability, $P_j(n:t,a,b)$, to observe exactly $n$ pre-spikes in the time interval $[0,t]$ and in the space interval $[a,b]$ (where $0<a<b<1$), and that no jump occurs in the time interval $[0,t]$, starting from $0$. The corresponding generating function in the Laplace domain $Z(s:\sigma,a,b)$, defined in Eq.~\eqref{eq:Zc}, can be obtained by plugging $C(\sigma)$, $D(\sigma)$ and $E(\sigma)$ into Eq.~\eqref{eq:functionZ}.
It follows that for any $s\in [0,1]$ fixed we have in the large $\gamma$ limit that
\begin{equation} 
\label{eq:Z_thm}
   \lim_{\gamma\rightarrow\infty} Z(s:\sigma,a,b)= \cfrac{1}{\sigma +W_{-,+}+W_{-,+} (1-s) \Big(\frac{1}{a}-\frac{1}{b}\Big)} \ .   
\end{equation}

The probability $S(t)$ of no jump occurring from $0$ to $1$ in the time interval $(0,t)$ is obtained by taking $s=1$ in the generating function in Eq.~\eqref{eq:Z_Rabi}, which yields $S(t)=e^{-W_{-,+} t}$. This implies that the first time to reach $1$ starting from $0$ is exponentially distributed with a mean equal to $1/W_{-,+}$. The probability $P_c(n : t, a, b)$ to observe exactly n
spikes in the time interval $[0, t]\times[a, b]$, \textit{given} that no
jump occurs in the time interval $[0, t]$, can be obtained by 
\begin{equation*}
P_c(n:t,a,b)=\frac{P_j(n:t,a,b)}{S(t)}\,,    
\end{equation*} 
which is given by the Poisson distribution in the large $\gamma$ limit (as announced in Eq.~\eqref{eq:spikes-stat}),
\begin{equation}
        \frac{(\lambda_{[a,b]} t)^ne^{-\lambda_{[a,b]}t}}{n!},
  \hspace{0.2cm}\text{where}\hspace{0.1cm}\lambda_{[a,b]} =\int_a^bdx\,\tfrac{W_{-,+}}{x^2} \ .    
\end{equation}
Moreover, similar to the Collapse-Unitary setup, we have that the probability to have no resetting to $0$ by starting from $1$ in a time interval of length $\Delta t$ is given for large $\gamma$ by
\begin{equation}
    \mu (\Delta t \vert 1) \approx \exp\left( - W_{+,-} \Delta t \right)\ ,
\end{equation}
so that the time to be resetting to $0$  when starting from $1$ is exponentially distributed with mean $1/W_{+,-}$.

\bigskip

\subsection{Study in the Collapse-Measurement setup}
To simplify notation, we denote ${\mathcal N}^{1, \gamma_1}$ by ${\mathcal N}^1$, ${\mathcal N}^{2, \gamma_2}$ by ${\mathcal N}^2$ and $q_t^\gamma$ by $q_t$. Eq.~\eqref{eq:sde-2-2-1-1} can then be rewritten as

\begin{equation}
\label{eq:SDEMain-cc}
dq_{t}=\Omega\left(q_{t}\right)dt-q_{t}\ d{\mathcal N}^{1}_{t}+(1-q_{t})\ d{\mathcal N}^{2}_t \ ,
\end{equation}
where the quadratic drift term $\Omega:=\Omega^{\gamma_1, \gamma_2}$ and the rate functions $\alpha$ of $\mathcal{N}^1_t$ and $\tilde{\alpha}$ of $\mathcal{N}^2_t$ are given by:
\begin{equation}
\begin{split}
 &\Omega(q) =(\gamma_2+q\gamma_1\eta_1)(1-q)-\gamma_2\eta_2(1-q)^2,\\
&\mathbb{E}\left(d{\mathcal N}^{1}_{t} \vert q_t \right) =\alpha(q_t)\,dt=\gamma_1\eta_1(1-q_t)dt \ , \\
&\mathbb{E}\left(d{\mathcal N}^{2}_{t}  \vert q_t \right) =\tilde{\alpha}(q_t)\,dt=\gamma_2\eta_2(1-q_t)dt \ .
\end{split}
\end{equation}
The roots of $\Omega(q)$ are the fixed points of the no-click dynamics, given by
\begin{equation}
q_-=\tfrac{\gamma_2(\eta_2-1)}{\gamma_2\eta_2+\gamma_1\eta_1} \ , \quad q_+=1 \ .
\end{equation}
Note that $q_-<0$ (provided $N_2$ is not measured with unit efficiency).

\medskip

The solution of the deterministic equation $d\dot{\hat q}_t=\Omega (\hat q_t) dt$ , starting from $q_0$ at time $t=0$, is given by 
\begin{align}
{\hat q}_t(q_0)&=\frac{(q_0-q_-)+q_-(1-q_0)e^{-(\gamma_1\eta_1+\gamma_2\eta_2)(1-q_-)t}}{(q_0-q_-)+(1-q_0)e^{-(\gamma_1\eta_1+\gamma_2\eta_2)(1-q_-)t}} \ .
\label{eq:colsol}
\end{align}
For $c \in [0,1]$, let $\tau_c$ be the time taken to evolve from $q_0=0$ to $q=c$. A straightforward inversion of Eq.~\eqref{eq:colsol} show that

\begin{equation}
\tau_c= \frac{1}{(\eta_1\gamma_1+\eta_2\gamma_2)(1-q_-)}\log\Bigg|\frac{(c-q_-)}{q_-(1-c)}\Bigg| \  .
\end{equation}

Let $\tau$ be the time taken by a deterministic trajectory to evolve from $q=0$ to $q=1-\epsilon$. 
For $\gamma_1\gg 1$,
\begin{equation*}
\tau\sim\frac{1}{\gamma_1\eta_1}\log\Bigg|\frac{\gamma_1\eta_1\left(1-\epsilon\right)}{\epsilon\gamma_2(\eta_2-1)}\Bigg|   \ . 
\end{equation*}

Similar to the Collapse-Unitary and Collapse-Thermal cases, we explicitly obtain the functions
\begin{equation}
\begin{split}
\label{eq:Alan014}
        \mu(t|q_0) &= e^{-\int_0^t ds (\alpha[{\hat q}_s(q_0)]+ \tilde{\alpha}[{\hat q}_s(q_0)])} \ , \\
        &=\frac{(q_0-q_-)+(1-q_0)e^{-(\gamma_1\eta_1+\gamma_2\eta_2)(1-q_-)t}}{1-q_-} \ ,  \\
     \mu(t|q_0)& \alpha [{\hat q}_t(q_0)]\\
     &=\gamma_1\eta_1(1-q_0)e^{-(\gamma_1\eta_1+\gamma_2\eta_2)(1-q_-)t} \ , 
\end{split}
\end{equation}
which we plug into the Eq.~\eqref{eq:CDE} to explicitly obtain the functions $C(\sigma)$, $D(\sigma)$ and $E(\sigma)$ and obtain their asymptotic behaviour for large $\gamma$ (See Appendix \ref{app:LaplaceTransform} for details).

\medskip 
Consider the probability, $P_j(n:t,a,b)$, to observe exactly $n$ pre-spikes in the time interval $[0,t]$ and in the space interval $[a,b]$ (where $0<a<b<1$), and that no jump occurs in the time interval $[0,t]$, starting from $0$. The corresponding generating function in the Laplace domain $Z(s:\sigma,a,b)$, defined in Eq.~\eqref{eq:Zc}, can be obtained by plugging $C(\sigma)$, $D(\sigma)$ and $E(\sigma)$ into Eq.~\eqref{eq:functionZ}. It follows that for any $s\in [0,1]$ fixed we have in the large $\gamma_1$ limit (for $\gamma_2$ fixed)
\begin{equation}
\lim_{\gamma_1\rightarrow\infty} Z(s:\sigma,a,b)= \cfrac{1}{\sigma +\gamma_2+(1-s)\gamma_2(1-\eta_2) \Big(\tfrac{1}{a}-\tfrac{1}{b}\Big)} \ .
    \label{eq:Z-cc}
\end{equation}

The probability $S(t)$ of no jump occurring from $0$ to $1$ in the time interval $(0,t)$ is obtained by taking $s=1$ in the generating function in Eq.~\eqref{eq:Z_Rabi}, which yields $S(t)=e^{-\gamma_2 t}$. This implies that the first time to reach $1$ starting from $0$ is exponentially distributed with a mean equal to $1/\gamma_2$ and that the statistics of the spikes given the no-jump condition is given by the Poisson distribution in the large $\gamma_1$ limit (as announced in Eq.~\eqref{eq:spikes-stat}),
\begin{equation}
        \frac{(\lambda_{[a,b]}t)^ne^{-\lambda_{[a,b]}t}}{n!},
  \hspace{0.2cm}\text{where}\hspace{0.1cm}\lambda_{[a,b]} =\int_a^bdx\,\tfrac{\gamma_2(1-\eta_2)}{x^2} \ .    
\end{equation}

Note that, in this setup, the probability to have resetting to $0$ by starting from $1$ vanishes and the flow term out of $1$ also vanishes. The mean time of first passage from $0$ to $1$ is $\gamma_2^{-1}$ and hence, spikes are typically not observed for $t\gg\gamma_2^{-1}$.

\bigskip

\section{Generalization: a general perspective on spikes in 1D piecewise deterministic Markov process}
\label{sec:conjecture}

Our interest in spikes processes was motivated by quantum physics and its analytical description was limited by various technicalities, see the discussion below. But we may enlarge the discussion to cover other physical situations of interest. 
\medskip

To this end, let us consider a piecewise $1$-dimensional Markov process $(q^\gamma_t)_{t \ge 0}$ evolving according to the SDE
\begin{equation}
\label{eq:generalresetting}
\dot q_t ^\gamma =  F(q_t^\gamma) + G(q_t^\gamma) \left( {\dot{\mathcal N}}_t^\gamma - \gamma H (q_t^\gamma) \right)\ ,
\end{equation}
where ${\mathcal N}^\gamma$ is a Poisson process satisfying
\begin{equation*}
\mathbb E (d \mathcal N_t^\gamma \vert q_t^\gamma) = \gamma H (q_t^\gamma) dt, \quad d{\mathcal N}_t^\gamma d{\mathcal N}_t^\gamma = d{\mathcal N}_t^\gamma \ .
\end{equation*}
We assume that $G(q)>0$ if $q\in\left]0,1\right[$ and that  
\begin{equation}
G(0)H(0)=0=G(1)H(1), \quad F(0) \ge 0, \quad F(1) \le 0 \ ,
\label{eq:hyp}
\end{equation}
so that the process $q^\gamma$ takes values in $[0,1]$. We conjecture that the graph $\{(t, q_t^{\gamma})\; ; \; t \ge 0\}$ of the process $q^\gamma$ converges as $\gamma \to \infty$ to the random time-space ``spike''  graph $\mathbb Q$, which can be described as follows.

\medskip 
Let $(\bar{q}_t)_{t \ge 0}$ be the continuous pure jump Markov process on $\{0,1\}$ with rate $F(0)$ (resp. $F(1)$) to jump from $0$ to $1$ (resp. from $1$ to $0$). The process ${\bar q}$ can be seen as the first rough description of $q^\gamma$ for large $\gamma$, but it misses a more refined structure of the limit of $q^\gamma$.

\medskip
``Decorate'' then the trajectory $\bar q$ with spikes, which are vertical intervals (in space) included in $[0,1]$,  distributed as \footnote{Here ${\bar q}_{t-}=, {\bar q}_{t-\varepsilon}$ is the left limit of $\bar q$ at time $t$. By convention we assume that the process $\bar q_t$ has continuous from the right trajectories.}
\begin{equation}
\label{eq:mathbbQ}
\mathbb Q_t =
\begin{cases}
&[0, M_t], \quad\text{if} \quad {\bar q}_{t-}= {\bar q}_t =0 \ ,\\
&[M_t, 1], \quad\text{if} \quad {\bar q}_{t-}= {\bar q}_t =1 \ ,\\
&[0,1], \hspace{0.03\linewidth}\quad\text{if} \quad {\bar q}_{t-} \ne {\bar q}_t \ ,
\end{cases}
\end{equation}
where, given the trajectory ${\bar q}$,  $(t,M_t)_{t \ge 0}$ is the time-space Poisson process on $[0,\infty) \times [0,1]$.  

\begin{figure}
    \centering
\includegraphics[width=\linewidth]{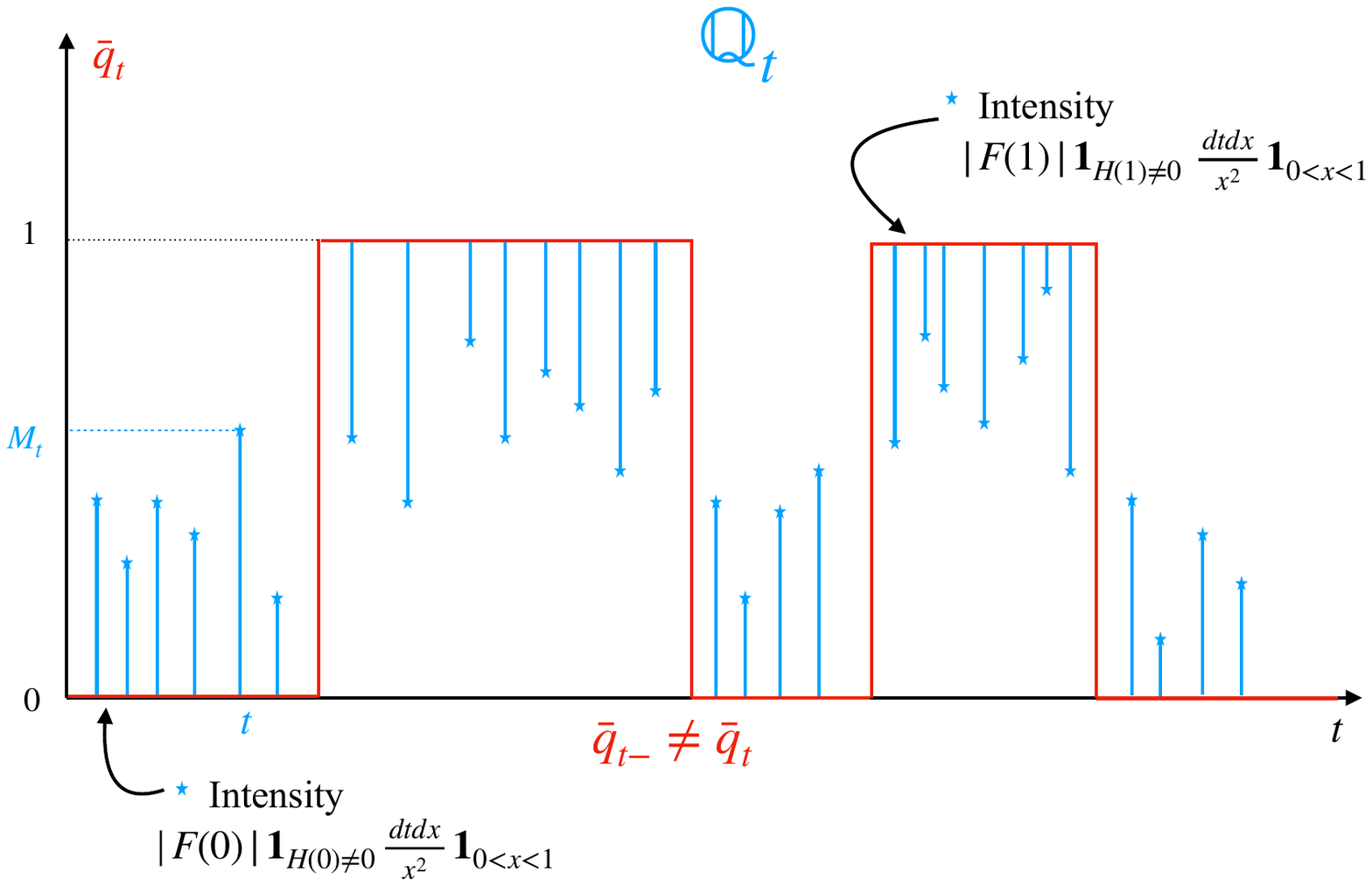}
    \caption{The spike process $\mathbb Q_t$. The value $M_t$ is the length of the spike at time $t$. The condition ${\bar q}_{t-}= {\bar q}_t =0$ (resp.  ${\bar q}_{t-}= {\bar q}_t =1$) corresponds to the case where the spike starts from $0$ (resp. $1$), while the condition ${\bar q}_{t-} \ne {\bar q}_t$ corresponds to a jump for the process $\bar q_t$.}  
    \label{fig:spikeprocessgraph}
\end{figure}

More precicely, the spikes from $0$ to $1$ have intensity $\vert F(0) \vert {\mathbf 1}_{H(0) \ne 0} \, \tfrac{dt \, dx}{x^2} \,{\mathbf 1}_{x \in [0,1]}$ and the spikes from $1$ to $0$ have intensity $\vert F(1) \vert {\mathbf 1}_{H(1) \ne 0} \, \tfrac{dt \, dx}{(1-x)^2} \,{\mathbf 1}_{x \in [0,1]}$. See Fig. \ref{fig:spikeprocessgraph}.

\medskip 
More explicitly, let $P_c [n\,:(s,t),a,b]$ be the probability of observing $n$ tips of pre-spikes belonging to $(a,b)$, conditioned on no jumps occurring during the time window $(s,t)$. Its large $\gamma$ limit corresponds precisely to a  Poisson distribution limit with parameter:
$
     \lambda_{\left[a,b\right]}^{{\bar q}} \, (t-s) \ , 
$
\begin{equation}
\label{eq:=conj}
\text{\rm{i.e.}} \quad \lim_{\gamma\rightarrow\infty}P_{c}\left[n:(s,t),a,b\right]=\tfrac{(t-s)^{n}}{n!}e^{-\lambda_{[a,b]}^{\bar q}(t-s)}\ ,
\end{equation}
with the intensity parameter given by
\begin{equation*}
        \lambda_{\left[a,b\right]}^{\bar q}=
\begin{cases}
        &\int_{a}^{b}dx\tfrac{\vert F (0) \vert \; {{\mathbf 1}_{H\left(0\right)\neq0}}}{x^{2}} \quad  \textrm{if } \quad {\bar q} \ \big\vert_{[s,t]}=0 \ , \\
        &\\
        &\int_{a}^{b}dx\tfrac{\vert F (1) \vert \; {\mathbf 1}_{H (1)\neq 0}}{(1-x)^{2}}  \quad  \textrm{if } \quad {\bar q} \ \big\vert_{[s,t]}=1 \ . 
\end{cases}
\end{equation*}

\bigskip

\begin{figure*}
    \centering
    \includegraphics[width=\linewidth]{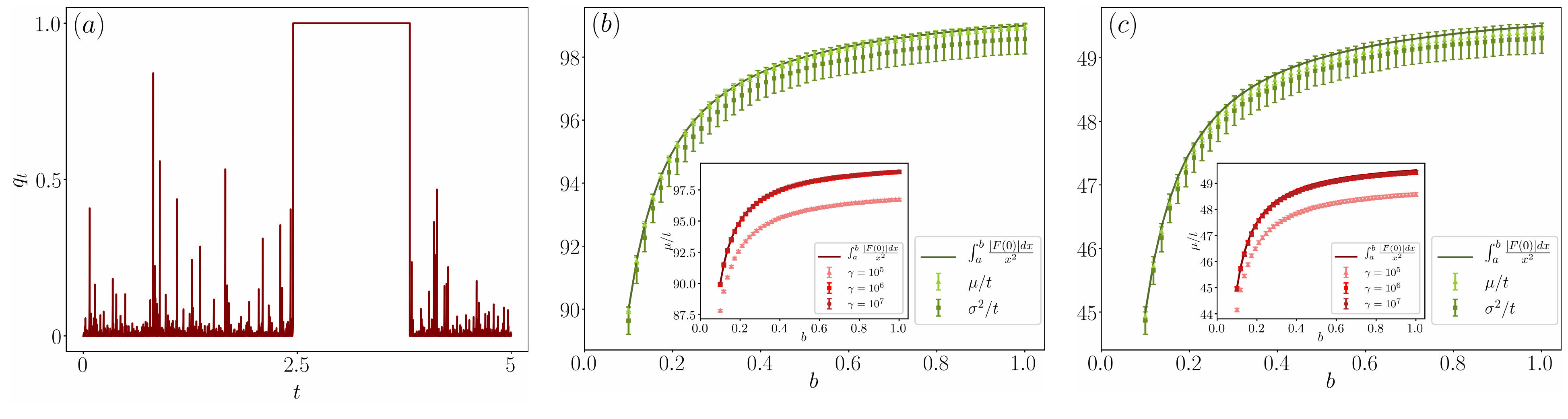}
    \caption{(a) The typical trajectory for the general 1D PDMP (Eq.~\eqref{eq:generalresetting}) with resetting to $q^*=0$ for $\gamma=10^4$. The jumps between $q_t=0$ and $q_t=1$ are interspersed with spikes from $q_t=0$ in the strong measurement limit. In $(b)$ and $(c)$, we plot the mean and variance of the distribution (conditioned on no-jumps) of the number of spikes (from $q_t=0$) per unit time in a space-time box $(0.01,b)\times (0,t)$ against the box-edge $b$ for (b) $F(q)=\cos\left(\tfrac{50q}{\pi}\right)$ and (c) $F(q)=(e^{-q}-0.5)$. This empirically verifies the conjecture that the spiking process is Poissonian with the parameter $\int_a^b\tfrac{|F(0)|dx}{x^2}$. The data-points were obtained for the parameters $t=0.1$ and $\gamma=10^7$ by averaging over $10^5$ realizations with $dt=6.25\times 10^{-11}$. The data in the insets show the convergence of the empirical mean/time to the conjectured value as $\gamma$ is increased.}
    \label{fig:general-spikes}
\end{figure*}

\bigskip 

\label{sec:technicalities}
Observe also that in this general context (not motivated by quantum physics), to adapt the analytical proofs developed in this paper, we would need: 
\begin{enumerate}
\item to have a resetting dynamics, i.e. $G(q)= q^*-q$, $q^* \in (0,1)$ \ ;
\item  to have a good understanding of the deterministic flow ${\hat q}^\gamma$ defined by $\dot {\hat q}_t^\gamma = F({\hat q}_t^\gamma) - \gamma G({\hat q}_t^\gamma) H ({\hat q}_t^\gamma)$ for large $\gamma$ \ .
\end{enumerate}

Note that the Collapse-Unitary case considered in  section~\ref{sec:mainresults} and \ref{sec:an-CUsetup} is not in the form of our  general set-up Eq.~\eqref{eq:generalresetting} on $[0,1]$.
To transform Eq.~\eqref{eq:sde-angle-1-1-1-1}  to an equation on the line as  inEq.~\eqref{eq:generalresetting}, we need to come back to the usual coordinate $q^\gamma$ in equation Eq.~\eqref{eq:sde-2}. 
 With the notations of Section \ref{sec:mainresults}, the plane $\varphi_{t}=\frac{\pi}{2}$ is stable, and then by restricting the dynamics to this plane, we find  the equation 
\begin{equation}
\label{eq:gen}
{\dot q}_t^\gamma = F_\gamma (q_t^\gamma) + G(q_t^\gamma) \left( {\dot{\mathcal N}}_t^\gamma - \gamma H (q_t^\gamma) \right)\ ,
\end{equation}
with 
\begin{equation*}
F_\gamma (q)=2 k_{\gamma}  \sqrt{ q(1-q)} , \quad G(q)= -q, \quad H(q)= \eta (1-q), 
\end{equation*}
and 
${\mathcal N}^\gamma$ is a Poisson process satisfying
\begin{equation}
\label{eq:CUP}
\mathbb E (d \mathcal N_t^\gamma \vert q_t^\gamma) = \gamma H (q_t^\gamma) dt, \quad d{\mathcal N}_t^\gamma d{\mathcal N}_t^\gamma = d{\mathcal N}_t^\gamma \ .
\end{equation}
At first sight, the equation Eq.~\eqref{eq:gen} is directly in the form Eq.~\eqref{eq:generalresetting} if $k_{\gamma}$ is $\gamma$ independent, but then the general previous conjecture gives that there are no jumps and no spikes because $F(0)=0=F(1)$ in Eq.~\eqref{eq:CUP} (Zeno effect). 
On the other hand, if  $k_{\gamma}=\sqrt{\omega \gamma}$ as we did in section \ref{sec:mainresults} and \ref{sec:an-CUsetup},  then $F_\gamma (q)=2 \sqrt{\gamma}  \sqrt{\omega q(1-q)}$ and the factor $\sqrt{\gamma}$ breaks the validity of the conjecture.

\begin{figure*}
 	\centering
 	\includegraphics[width=\linewidth]{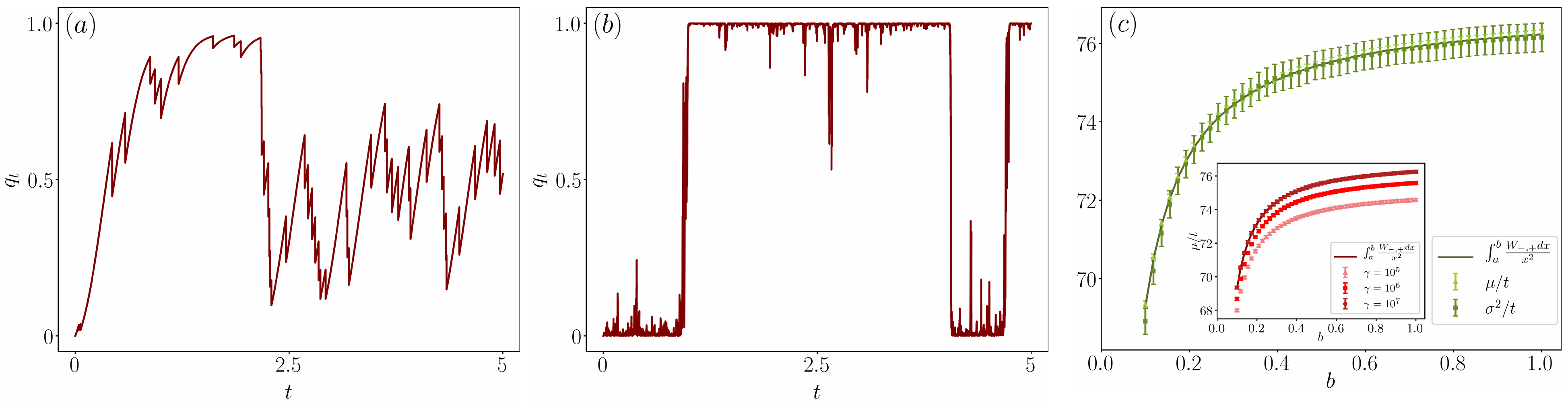}

\caption{The typical trajectory for (a) $\gamma=10^2$ (b) $\gamma=10^4$ for the Collapse-Thermal set-up for a general measurement operator for the parameters $|n_+|^2=0.2$, $|n_-|^2=0.4$, $W_{-,+}=0.77$, $W_{+,-}=0.23$. Here, (b) shows that the jumps between $q_t=0$ and $q_t=1$ are interspersed with spikes from both $q_t=0$ and $q_t=1$ in the strong measurement limit, even though the dynamics in this case does not reset to a fixed point. In $(c)$, we plot the mean and variance of the distribution (conditioned on no-jumps) of the number of spikes (from $q_t=0$) per unit time in a space-time box $(0.01,b)\times (0,t)$ against the box-edge $b$ for the same set of parameters. This empirically verifies the conjecture that the spiking process is Poissonian with the parameter $\int_a^b\tfrac{W_{-,+}dx}{x^2}$. The data-points were obtained for the parameters $t=0.1$ and $\gamma=10^7$ by averaging over $10^5$ realizations with $dt= 10^{-9}$. The data in the inset shows the convergence of the empirical mean/time to the conjectured value as $\gamma$ is increased.}
\label{fig:traj-thermal-2}
 \end{figure*}

 Below we first give, in Sec.~\ref{subsec:ex1}  some non-quantum motivated examples where we numerically verify 
the above conjecture. In Sec.~\ref{subsec:ex2} we discuss a non-trivial quantum example where we expect the conjecture to be true but our analytical approach based on resetting cannot be applied. Finally in Sec.~\ref{subsec:ex3} we discuss a generalization of Eq.~\eqref{eq:generalresetting} in the context of collapse-measurement set-up.

\subsection{Example 1 with resetting: Non-Quantum SDEs}
\label{subsec:ex1}
In the general Markovian 1D piecewise deterministic process  given by Eq.~\eqref{eq:generalresetting}, we numerically study the emergence of spikes for large $\gamma$.
In the context of resetting dynamics to a point $q^*$,  we require $G(q)=q^*-q$. If $q^*\neq 0$ and $q^*\neq1$, the hypothesis Eq.~\eqref{eq:hyp} will only be satisfied if $H(0)=H(1)=0$, in which case the intensity of the spiking process vanishes. Thus, to observe spikes, we take $q^*=0$ without loss of generality. In addition, to satisfy Eq.~\eqref{eq:hyp}, we need then to take $H(1)=0$, so we consider $H(q)=(1-q)\chi(q)$, where $\chi(q)$ is any finite function such that $\chi(0) \neq 0$. Then the conjecture Eq.~\eqref{eq:=conj} gives that  $P_c [n\,:(s,t),a,b]$ converge at large $\gamma$ to a  Poisson distribution limit Eq.~\eqref{eq:=conj} with intensity parameter given by (one-sided spikes)
\begin{equation*}
        \lambda_{\left[a,b\right]}^{\bar q}=
\begin{cases}
        &\int_{a}^{b}dx\tfrac{\vert F(0) \vert \; }{x^{2}} \quad  \textrm{if } \quad {\bar q} \ \big\vert_{[s,t]}=0 \ , \\
        &\\
        & 0  \quad  \textrm{if } \quad {\bar q} \ \big\vert_{[s,t]}=1 \ . 
\end{cases}
\end{equation*}
In Fig. \ref{fig:general-spikes},  we numerically verify the conjecture by taking $\chi(q)$ to be a fourth-order polynomial with arbitrary coefficients, for two very different forms of $F(q)$:\\ (i) $F(q)=\cos\left(\tfrac{50q}{\pi}\right)$, a highly oscillatory function and\\ (ii) $F(q)=(e^{-q}-0.5)$, a strictly monotonic function.

\subsection{Example 2 without resetting:  Collapse-Thermal setup with general QND (diagonal) measurement $N_1$ operator}
\label{subsec:ex2}

Consider the Collapse-Thermal setup for a qubit with a general diagonal measurement operator (intensity is denoted by $\gamma$ and efficiency by $\eta$) 
$$N_1=\left(\begin{array}{cc}
n_{+} & 0\\
0 & n_{-}
\end{array}\right) \ .$$

By using the computations performed in Appendix \ref{app:1}, we easily get that, with the usual parameterization of the density matrix $\rho_{t}^\gamma$ given in Eq.~\eqref{eq:parameter1},

\begin{equation}
\label{eq:EMS-3-1-2-1-1-1-1-1}
\begin{split}
&\dot q_{t}^{\gamma} =(W_{+,-}+W_{-,+}) \big(\tfrac{W_{-,+}}{W_{+,-}+W_{-,+}}-q_{t}^{\gamma}\big) \\
& +\tfrac{\vert n_{+}\vert ^{2}-\vert n_{-}\vert ^{2}}{(\vert n_{+}\vert ^{2}-\vert n_{-}\vert ^{2} )q_{t}^{\gamma}+\vert n_{-}\vert ^{2}}\, (1-q_{t}^{\gamma})q_{t}^{\gamma}\, \left(\dot {\mathcal N}_{t}^{\gamma} - {\mathbb E}\left(\dot{\mathcal N}_{t}^\gamma \vert q_{t}^{\gamma}\right) \right)\ ,  
\end{split}
\end{equation}
with $d{\mathcal N}_{t}^{\gamma} d{\mathcal N}_{t}^{\gamma}=d{\mathcal N}_{t}^{\gamma}$
\begin{equation}
\label{eq:Poi-1-1-1-1-1}
\mathbb{E}\left(d{\mathcal N}^\gamma_{t}\vert q_{t}^{\gamma}\right)=\gamma\eta \left[ (\vert n_{+}\vert^{2}-\vert n_{-}\vert^{2})\, q_{t}^{\gamma}+\vert n_{-}\vert^{2} \right]dt \ .
\end{equation}
Observe that Eq.~\eqref{eq:EMS-3-1-2-1-1-1-1-1} takes the form of Eq.~\eqref{eq:generalresetting} with 
\begin{equation}
\begin{split}
F(q) & = (W_{+,-}+W_{-,+}) \big(\tfrac{W_{-,+}}{W_{+,-}+W_{-,+}}-q_{t}^{\gamma}\big) \ , \\
G(q) &= \tfrac{\vert n_{+}\vert ^{2}-\vert n_{-}\vert^{2}}{(\vert n_{+}\vert^{2}-\vert n_{-}\vert^{2} )q +\vert n_{-}\vert^{2}}\, (1-q )q \ , \\
H(q)&= \eta \left[ (\vert n_{+}\vert^{2}-\vert n_{-}\vert^{2})\, q +\vert n_{-}\vert^{2} \right]\ .
\end{split}
\end{equation}

\bigskip

In particular, for the model considered in Eq.~\eqref{eq:EMS-3-1-2-1-1-1-2}, i.e. $n_+=0$, $n_-=1$ (resetting dynamics), we have that our conjecture is satisfied with spikes only from $0$ and spikes statistics described by the spike process $\mathbb Q$ defined in Eq.~\eqref{eq:mathbbQ}.

\bigskip 
But our conjecture claims that, moreover, in general (even without resetting dynamics), there are spikes starting from $0$ and from $1$. This is the case if $F(0)H(0) = \eta W_{-,+} |n_-|^2 \ne 0$ (spikes from $0$ to $1$) and $F(1)H(1)= \eta W_{+,-} \vert n_+\vert^2 \ne 0$ (spikes from $1$ to $0$). This is confirmed by numerical simulations (see Fig. \ref{fig:traj-thermal-2}).

\bigskip

Observe moreover that if we consider the measurement operator $N_1=\sigma_{z}$ (hence $\vert n_-\vert^2= \vert n_+\vert^2$), as it was considered in the Gaussian case \cite{Bauer,Tilloy,Bauer2018,BCCNP2023}, we get a purely deterministic equation without jumps or spikes. This is very different from the studies cited previously in the Gaussian noise context.

\subsection{Example 3 with resetting: Collapse-Measurement setup for QND resetting $N_1=\vert -\rangle \langle -\vert$ but general $N_2$ operator}
\label{subsec:ex3}

\medskip

 To generalize the Collapse-Measurement setup given by equation Eq.~\eqref{eq:sde-2-2-1-1}, we can substitute the damping-spontaneous emission operator $N_2$, defined in Eq.~\eqref{eq:N2-007}, by a general (not necessarily diagonal) measurement operator  $N_2$ (in order to break the QND assumption).  
Then, we can check that, generically, the equation for $q^\gamma$ is not autonomous. More exactly, the non-commutativity of $N_2$ w.r.t. $N_1$ (to break the QND assumption) and the fact that $q^\gamma$ has an autonomous evolution, are equivalent to taking $N_2$ in the form 
\begin{equation}
\label{eq:N2gen}
N_{2}=\left(\begin{array}{cc}  
0 & a\\
b & 0
\end{array}  \right), \quad ab \ne 0  \ . 
\end{equation}

\medskip
The requirement that $q^\gamma$ has an autonomous evolution seems to be a necessary condition to adapt our analytical proof (technical assumption).

\medskip 
Observe, however that, even in this case, the Collapse-Measurement setup does not fall directly in our conjecture case Eq.~\eqref{eq:generalresetting} because we have then two Poisson noises involved (but the second one is weaker than the first one since $\gamma_2$ is of order one while $\gamma_1$ is very large). More precisely, the equation for $q^\gamma$ is given by (Appendix \ref{sec:Appendix-:new})
\begin{equation}
\label{eq:general-collapse-measuerment}
\begin{split}
{\dot q}_{t}^{\gamma}& = F(q_t^\gamma) + G_1 (q_t^\gamma) \left( {\dot{\mathcal N}}^{1,\gamma}_t - \gamma_1 H_1 (q_t^\gamma)  \right) + G_2 (q_t^\gamma) {\dot{\mathcal N}}^{2,\gamma}_t  \ ,
\end{split}
\end{equation}
with 
\begin{equation*}
\begin{split}
F (q)&= \gamma_2 \left\{ \vert a \vert^2 - (\vert a \vert^2 +\vert b \vert^2 ) q \right\} \left\{1 - \eta_2  G_2 (q)  \right\}  \ ,
\\
 G_2 (q)&=\tfrac{\vert a \vert^2 (1-q) }{\vert a \vert^2 (1-q) + \vert b \vert^2  q  } - q \ , \\
 H_2 (q) &=  \eta_2  \left( \vert a \vert^2 (1-q) + \vert b \vert^2 q  \right) \ ,
\\
 G_1 (q) &= - q \ , \\
 H_1 (q)& = \eta_1 (1-q) \ , 
 \end{split}
\end{equation*}
and the Poisson noises satisfy
\begin{equation*}
\begin {split}
\mathbb{E}\left( d{\mathcal N}_{t}^{k,\gamma}\vert q_{t}^{\gamma}\right)=H_k (q_t^\gamma) dt \ , \\
 d{\mathcal N}_{t}^{k,\gamma} d{\mathcal N}_{t}^{\ell,\gamma}=\delta_{k\ell} \ d{\mathcal N}_{t}^{k,\gamma} \ .
\end{split}
\end{equation*}

\medskip

Hence, a stronger conjecture of the one above is that, if we have an evolution equation given by Eq.~\eqref{eq:general-collapse-measuerment} with hypothesis Eq.~\eqref{eq:hyp}, then in the large limit $\gamma_1 \to \infty$, $\gamma_2$ of order $1$, the graph of $q^\gamma$ should converge to a spiking process $\mathbb Q$, given by Eq.~\eqref{eq:mathbbQ} with $H$ replaced there by the dominant term $H_1$. Since $F(1)H_1(1)=0$, we expect only to have no spikes from $1$ (apart from the  degenerate case $\gamma_2 \vert a \vert^2 \left(1 - \eta_2  \right) =0$, see below, we will always have spikes from $0$).

\medskip

In the present paper, we considered only $N_2$ equal to the damping-spontaneous emission operator appearing in Eq.~\eqref{eq:N2-007}, which corresponds to the case $a=1$ and $b=0$, and then the previous conjecture about the spike statistics is in agreement with the third relation in Eq.~\eqref{eq:spikeintense} because $F(0)=\gamma_2 \left(1-\eta_2\right)$. Another standard rank one choice is to take $N_2$ equal to the spontaneous absorption operator given by
\[
N_{2}=\left(\begin{array}{cc}
0 & 0\\
1 & 0
\end{array}  \right) = \vert - \rangle \langle + \vert \ . 
\]  
In the latter case ($a=0, b=1$) we have then a competition between a strong resetting dynamics and a weak resetting dynamics towards the same $\vert - \rangle \langle - \vert$. Our conjecture implies thus that in this case, we do not have spikes but only jumps.

\bigskip

\medskip

\section{Conclusion}
\label{sec:conclusion}

In this work, we discussed finite-dimensional quantum trajectories driven by Poissonian noises, see Eq.~\eqref{eq:EMS-3}, in a strong measurement limit and focused our studies to a qubit system in three different setups where the Quantum-Non-Demolition condition for the strong measurement is broken by a unitary evolution, a thermal bath or a second measurement, see Sec.\ref{sec:precisesetup}. Our analytical studies of Sec. \ref{sec:derivations} establish that the quantum trajectory investigated behaves like a pure jump Markov process in the strong measurement limit. However, this rough picture can be refined and it appears in fact that this quantum jump Markov process is in fact decorated by vertical lines (called spikes). The first analytical description of spikes was done recently by Bauer-Bernard-Tilloy~\cite{Bauer,Tilloy,Bauer2018}, and motivated some others works~\cite{Kolb,BCCNP2023}, but in the context of quantum trajectories of a qubit driven by a Gaussian noise. 

This paper is therefore the first to investigate this question for quantum trajectories driven by Poissonian noises. 

Observe that in all the setups we studied analytically, we observe a kind of universal law for the distribution of spikes, see Eq.~\eqref{eq:spikes-stat}, valid for Gaussian or Poissonian noises. Numerical experiments confirm our analytical results. 

A fundamental property used to perform our analytical investigation in Sec. \ref{sec:derivations} is that the evolution of the quantum trajectory can be reduced to the evolution of a one-dimensional piecewise Markovian deterministic process with resetting. Hence, apart from its interest in this quantum framework, this problem could be subject of investigation in the resetting literature. Moreover, it seems that the existence of spikes is more generic and they can appear in other piecewise Markovian deterministic processes which do not have a resetting property, see e.g. Fig. \ref{fig:traj-thermal-2}. In Sec. \ref{sec:conjecture} we provide a general conjecture for the distribution of the spikes in a general one-dimensional framework. 
\medskip

Quantum trajectories are now routinely realized in experiments~\cite{Guerlin,Gleyzes,Sayrin,Murch2013,Murch2013-2,Weber,Lange}, but the question of whether spikes can be experimentally detected remains open. A few remarks on how this work addresses this question are in order. Firstly, spikes emerge in the $\gamma \to \infty$ limit, but their statistical properties can be observed for large but finite $\gamma$, as evidenced by our numerical results. These are obtained by the counting statistics  of "pre-spikes". Secondly, for the quantum trajectories of the kind considered in this work, the dynamics consists of a deterministic flow with intermittent resetting to a particular state. This dynamics does not merely arise from a toy model, but has its basis in modern day experiments \cite{minev2019catch}. Finally, one can truly claim the experimental detection of spikes only if one can infer the statistics of the spikes (or more precisely, the pre-spikes) \textit{without relying on the stochastic master equation}. This would take the observation of spikes out of the realm of simulations and into a laboratory. To achieve this, one can consider the quantum trajectory that was experimentally realized in \cite{minev2019catch}. Although that particular quantum trajectory had Gaussian noise, the experiment demonstrated highly efficient detection of clicks (transitions to the ground state) and moreover, by performing quantum state tomography over an ensemble of no-click trajectories, the state was shown to follow a fixed path through the Hilbert space during the no-click evolution. Using the knowledge of this path and the click sequence records, it is in principle possible to infer the statistics of the heights of the pre-spikes. However, there are a few subtleties associated with this proposal, one of them being that the system in \cite{minev2019catch} was essentially a qutrit. We expect that spikes can occur in higher dimensional quantum systems, but it is not known whether they have the same  power-law statistics as the qubit case. These are questions that we hope to address in a future work.

\acknowledgments
We thank Michel Bauer, Reda Chhaibi, Clement Pellegrini for useful discussions. This work was supported by the projects RETENU ANR-20-CE40 of the French National Research Agency (ANR). 
AS and AD  acknowledge financial support of the Department of Atomic Energy, Government of India, under Project Identification No. RTI 4001. AK acknowledges financial support for project NDFluc U-AGR-7239-00-C from the Luxembourg National Research Fund, Fonds National de la Recherche.
 AD acknowledges the J.C. Bose Fellowship (JCB/2022/000014) of the Science and Engineering Research Board of the Department of Science and Technology, Government of India. 
AD and AK thanks Laboratoire J A  Dieudonn\'e and INPHYNI, Nice for supporting a visit which initiated this project. AD thanks the National Research University Higher School of Economics, Moscow for supporting a visit. RC thanks ICTS-TIFR for supporting a visit for the completion of the project.

\appendix

\section{Measurement part of the SME Eq.~\eqref{eq:EMS-3} for qubit with a general diagonal measurement operator}
\label{app:1}

We perform here computations in a more general framework where the measurement operator $N$ is still diagonal 
\begin{equation*}
	N=\left(\protect\begin{array}{cc}
		n_{+} & 0\protect\\
		0 & n_{-}
		\protect\end{array}\right) 
\end{equation*}
but not necessary of rank one like in Eq.~\eqref{eq:operatorN}.

We consider hence only the dynamics generated by the measurement operator $N$ with intensity $\gamma>0$ and efficiency $\eta$:
\begin{equation}
	\label{eq:measu-N-general}
	\dot\rho_t^\gamma=\gamma L_{N }\left[ \rho_{t}^{\gamma}\right]+M_{N }\left[ \rho_{t}^{\gamma}\right]\left[ \dot{\mathcal{N}}_{t}^{\gamma} -\gamma \eta {\rm{Tr}} \left(N \rho_{t}^{\gamma} N^{\dagger}\right)\right]
\end{equation}
with the noise ${\mathcal N}_t^{\gamma}$ satisfying $d\mathcal{N}_{t}^{\gamma} \in\{0,1\}$, $d\mathcal{N}_{t}^{\gamma}d\mathcal{N}_{t}^{\gamma}=d\mathcal{N}_{t}^{\gamma }$, and 
\begin{equation}
	\label{eq:Poi-2}
	\mathbb{E}\left( d\mathcal{N}_{t}^{\gamma}\vert \rho_{t}^{\gamma}\right)=\eta\gamma \  {\rm{Tr}} \left(N \rho_{t}^\gamma N^{\dagger}\right)dt  \ .
\end{equation}

We use the usual parameterisation of the density matrix $\rho_{t}^\gamma$ given in Eq.~\eqref{eq:parameter1}. Then Eq.~\eqref{eq:measu-N-general} takes the form of a three dimensional (because even if $q^\gamma$ is real, $u^\gamma$ is complex) piecewise deterministic Markov process
\begin{equation}
	\label{eq:sde-2-1-2}
	\begin{split}
		\dot q_{t}^{\gamma}& =\alpha (q_t^\gamma) \ \left( \dot{\mathcal N}_{t}^{\gamma} -\mathbb{E}\left(\dot{\mathcal N}_{t}^{\gamma}\vert q_{t}^{\gamma}\right)\right)\\
		\dot u_{t}^{\gamma} & =-\frac{\gamma}{2} ( \vert n_{+}\vert ^{2}+\vert n_{-}\vert ^{2}-2\overline{n_{+}}n_{-}) u_{t}^{\gamma}\\
		&+ \beta(q_t^\gamma) \,   u_{t}^{\gamma}\left( \dot{\mathcal N}_{t}^{\gamma} -\mathbb{E}\left(\dot{\mathcal N}_{t}^{\gamma}\vert q_{t}^{\gamma}\right) \right)
	\end{split}
\end{equation}
with
\begin{equation}
	\label{eq:alpha-beta}
	\begin{split}
		\alpha (q_t^\gamma) &= \frac{ (\vert n_{+}\vert ^{2}-\vert n_{-}\vert ^{2}) q_{t}^{\gamma} (1-q_{t}^{\gamma})}{\left(\vert n_{+}\vert ^{2}-\vert n_{-}\vert ^{2}\right)q_{t}^{\gamma}+\vert n_{-}\vert ^{2}} \ , \\ 
		\beta (q_t^\gamma) & = \frac{\overline{n_{+}}n_{-}}{\left(\vert n_{+}\vert ^{2}-\vert n_{-}\vert ^{2}\right)q_{t}^{\gamma}+\vert n_{-}\vert ^{2}}-1 \ , 
	\end{split}
\end{equation}
and
\[
\begin{split}
	&\mathbb{E}\left( d{\mathcal N}_{t}^{\gamma}\vert q_{t}^{\gamma}\right)=\gamma\eta \{ (\vert n_{+}\vert^{2}-\vert n_{-}\vert^{2})q_{t}^{\gamma}+\vert n_{-}\vert^{2}\} \, dt \ , \\
	&d{\mathcal N}_{t}^{\gamma}d{\mathcal N}_{t}^{\gamma}=d{\mathcal N}_{t}^{\gamma} \ .
\end{split}
\]

Observe that the diagonal dynamics is autonomous.

\medskip

\subsection{Resetting dynamics or not}
The transition rate of the Poisson noise in Eq.~\eqref{eq:measu-N-general}  are given by 
\begin{equation}
	\begin{split}
		\label{eq:transitionrates1}
		W\left(\left(q,p\right)\rightarrow,\left(q',p'\right)\right) & =\gamma\eta ( (\vert n_{+}\vert^{2}-\vert n_{-}\vert^{2})q+\vert n_{-}\vert^{2}) \\
		&\times \delta\left(q'-\tfrac{\vert n_{+}\vert^{2}q}{\left(\vert n_{+}\vert^{2}-\vert n_{-}\vert^{2}\right)q+\vert n_{-}\vert^{2}}\right)  \\
		& \times\delta\left(p'-\tfrac{\overline{n_{+}}n_{-}p}{\left(\vert n_{+}\vert ^{2}-\vert n_{-}\vert ^{2}\right)q+\vert n_{-}\vert ^{2}}\right) \ .
	\end{split}
\end{equation}

Observe that these jumps correspond to a resetting dynamics towards a point if and only if for any $q,p$ we have that
\[
\tfrac{\vert n_{+}\vert ^{2}q}{\left(\vert n_{+}\vert^{2}-\vert n_{-}\vert^{2}\right)q+\vert n_{-}\vert^{2}}=k_{1}, \quad
\tfrac{\overline{n_{+}}n_{-}p}{\left(\vert n_{+}\vert^{2}-\vert n_{-}\vert^{2}\right)q+\vert n_{-}\vert^{2}}=k_{2} \ ,
\]
where $k_1, k_2$ are two arbitrary fixed constants, i.e. independent of $q,p$. We show easily that there is only two possibilities to have that. The first one  is that
\begin{equation*}
	k_{1}=0=k_2, \quad n_{+}=0, \quad i.e. \quad   N=\left(\begin{array}{cc}
		0 & 0\\
		0 & 1
	\end{array}\right) =N_1
\end{equation*}
and the second one is 
\begin{equation*}
	k_{1}=1, \quad k_{2}=0, \quad n_{-}=0, \quad i.e. \quad  N=\left(\begin{array}{cc}
		1 & 0\\
		0 & 0
	\end{array}\right)  \ . 
\end{equation*}

\subsection{Proof of Eq.~\eqref{eq:sde-2-1-2}}

The proof is based on elementary algebraic computations. We have that the Lindbladian term is given by

\begin{align*}
	& L_{N}\left[\rho_{t}^\gamma\right] =N\rho_{t}^\gamma N^{\dagger}-\frac{1}{2}N^{\dagger}N\rho_{t}^\gamma-\frac{1}{2}\rho_{t}^\gamma N^{\dagger}N\\
	& =\left(\begin{array}{cc}
		0 & -\frac{\vert n_{+}\vert ^{2}+\vert n_{-}\vert ^{2}-2n_{+}\overline{n_{-}}}{2}\text{\ensuremath{{\overline u}_t^{\gamma}}}\\
		-\frac{\vert n_{+}\vert ^{2}+\vert n_{-}\vert ^{2}-2\overline{n_{+}}n_{-}}{2}u_{t}^{\gamma} & 0
	\end{array}\right) \ .
\end{align*}
The  ''stochastic innovation'' term is given by 
\begin{align*}
	\label{eq:M-1-1}
	& M_{N}\left[\rho_{t}^\gamma\right] =\frac{N\rho_{t}^\gamma N^{\dagger}}{{\rm{Tr}}\left(N\rho_{t}^\gamma N^{\dagger}\right)}-\rho_{t}^\gamma
	=\left(\begin{array}{cc}
		\alpha (q_t^\gamma)  &  {\bar \beta} (q_t^\gamma)  \,  {\overline u}_t^{\gamma}\\
		\beta (q_t^\gamma) \,  u_{t}^{\gamma} & - \alpha (q_t^\gamma)
	\end{array}\right)
\end{align*}
where $\alpha, \beta$ have been defined in Eq.~\eqref{eq:alpha-beta}.

\section{Proof of Eq.~\eqref{eq:general-collapse-measuerment} and of  Eq.~\eqref{eq:sde-2-2-1-1}}
\label{sec:Appendix-:new}

We consider Eq.~\eqref{eq:N2gen}

\begin{equation}
	N_{2}=\left(\begin{array}{cc}  
		0 & a\\
		b & 0
	\end{array}  \right), \quad ab \ne 0  \ . 
\end{equation}

Observe  that
\begin{align*}
	L_{N_{2}}&\left[\rho_{t}^\gamma \right]  =N_{2}\rho_{t}^\gamma N_{2}^{\dagger}-\frac{1}{2}N_{2}^{\dagger}N_{2}\rho_{t}^\gamma-\frac{1}{2}\rho_{t}^\gamma N_{2}^{\dagger}N_{2}\\
	& = \left(\begin{array}{cc}
		\left|a\right|^{2}\left(1-q_{t}^{\gamma}\right)-\left|b\right|^{2}q_{t}^{\gamma} & a\overline{b}u_{t}^{\gamma}-\frac{\left|b\right|^{2}+\left|a\right|^{2}}{2}\overline{u_{t}^{\gamma}}\\
		\overline{a}b\overline{u_{t}^{\gamma}}-\frac{\left|b\right|^{2}+\left|a\right|^{2}}{2}u_{t}^{\gamma} & \left|b\right|^{2}q_{t}^{\gamma}-\left|a\right|^{2}\left(1-q_{t}^{\gamma}\right)
	\end{array}\right)
\end{align*}

and

\begin{align*}
	M_{N_{2}}&\left[\rho_{t}^{\gamma}\right]  =\tfrac{N_{2} \rho_{t}^\gamma N_{2}^{\dagger}}{{\rm{Tr}} \left(N_{2} \rho_{t}^\gamma N_{2}^{\dagger}\right)}-\rho_{t}^\gamma \\ =& \left(\begin{array}{cc}
		\frac{\left|a\right|^{2}\left(1-q_{t}^{\gamma}\right)}{\left|a\right|^{2}\left(1-q_{t}^{\gamma}\right)+\left|b\right|^{2}q_{t}^{\gamma}}-q_{t}^{\gamma} & \frac{a\overline{b}u_{t}^{\gamma}}{\left|a\right|^{2}\left(1-q_{t}^{\gamma}\right)+\left|b\right|^{2}q_{t}^{\gamma}}-\overline{u_{t}^{\gamma}}\\
		\frac{\overline{a}b\overline{u_{t}^{\gamma}}}{\left|a\right|^{2}\left(1-q_{t}^{\gamma}\right)+\left|b\right|^{2}q_{t}^{\gamma}}-u_{t}^{\gamma} & \frac{\left|b\right|^{2}q_{t}^{\gamma}}{\left|a\right|^{2}\left(1-q_{t}^{\gamma}\right)+\left|b\right|^{2}q_{t}^{\gamma}}-1+q_{t}^{\gamma}
	\end{array}\right)\ .
\end{align*}

\medskip

Then, with the equation Eq.~\eqref{eq:sde-2-1-2-1-10} for the  $N_1$-measurement part,  we obtain finally the equation Eq.~\eqref{eq:general-collapse-measuerment}. 
In the particular case of spontaneous emission Eq.~\eqref{eq:N2-007}:
\begin{equation}
	N_{2}=\left(\begin{array}{cc}
		0 & 1\\
		0 & 0
	\end{array}  \right) = \vert + \rangle \langle - \vert \
	,\end{equation} 
the equation Eq.~\eqref{eq:sde-2-1-2-1-10} becomes the equation Eq.~\eqref{eq:sde-2-2-1-1}.

\section{Proof of Eq.~\eqref{eq:sde-angle-1-1-1}}
\label{sec:Appendix-:-Prove}

We consider the SME Eq.~\eqref{eq:EMS-3} in the Collapse-Unitary setup Eq.~\eqref{eq:Rabi} (see Eq.~\eqref{eq:sde-2}  with the usual parameterisation of the density matrix $\rho_{t}^\gamma$ given in Eq.~\eqref{eq:parameter1}). 
\medskip

To simplify notation, we denote now $\gamma_1$ by $\gamma$ by $\eta$,  ${\mathcal N}_t^{\gamma, 1}$ by $\mathcal N_t^\gamma$.

\medskip

Parameterising now the density matrix in Bloch coordinates (Eq.~\eqref{eq:QIS-1}), we prove below that 
\begin{equation}
	\label{eq:Eqn-r}
	\begin{split}
		\frac{d}{dt} (r_{t}^{\gamma})^2 & =\gamma\left(\eta-1\right)2(1-q_{t}^{\gamma})\left( (r_{t}^{\gamma})^2+2q_{t}^{\gamma}-1\right) \\
		& +\left(1-(r_{t}^{\gamma})^2 \right)\left[\gamma\left(2q_{t}^{\gamma}-1\right)+\dot {\mathcal N}_{t}^{\gamma}\right].
	\end{split}
\end{equation}
Then, we can conclude that if the measurement is totally efficient, i.e. $\eta=1$, then the Bloch sphere $r=1$ (pure states) is stable under the dynamics. We will restrict us to this case now, i.e. pure state dynamics ({ $r_{t}^{\gamma}=1$)} and  totally efficient $\eta=1$.

\medskip

The Bloch sphere of pure states can be parameterised with the angles $\theta_t^{\gamma} \in (-\pi,\pi]$ , $\varphi_t^{\gamma} \in [0,\pi]$, see Eq.~\eqref{eq:QIS-1}:
\begin{equation}
	\label{eq:CV}
	q_{t}^{\gamma}=\cos^2 \left(\tfrac{\theta_t^{\gamma} }{2}\right), \quad 
	u_{t}^{\gamma}=\tfrac{\left(\sin\theta_t^{\gamma} \right)}{2}\, e^{i\varphi_t^{\gamma}}  \ .
\end{equation}
We then look for an SDE for $\left(\theta_t^{\gamma} ,\varphi_t^{\gamma} \right)$
which gives a solution to the sde Eq.~\eqref{eq:sde-2}. We claim that if $(\theta_t^\gamma, \varphi_t^\gamma)$ satisfies the sde 
\begin{equation}
	\label{eq:sde-angle-1-1}
	\begin{cases}
		\dot \theta_t^{\gamma} =\left(-2k_{\gamma} \sin (\varphi_t^\gamma)  -\frac{\gamma}{2}\sin\left(\theta_t^{\gamma} \right)\right)+\left(\pi-\theta_t^{\gamma} \right) \dot {\mathcal N}_{t}^{\gamma} \ , \\
		\dot \varphi_t^{\gamma} =-2k_{\gamma}\frac{\cos\left(\theta_t^{\gamma}\right)}{\sin\left(\theta_t^{\gamma}\right)}\cos\left(\varphi_t^{\gamma}\right)
	\end{cases}
\end{equation}
with 
\begin{equation*}
	\mathbb{E}\left( d{\mathcal N}_{t}^{\gamma}\vert \theta_t^{\gamma} \right)=\gamma\left(\sin\tfrac{\theta_t^{\gamma} }{2}\right)^{2}dt, \quad 
	d{\mathcal N}_{t}^{\gamma}d{\mathcal N}_{t}^{\gamma}=d{\mathcal N}_{t}^{\gamma} \ ,
\end{equation*}
then $(q_t^\gamma, u_t^\gamma)$ defined through Eq.~\eqref{eq:CV} satisfied the sde Eq.~\eqref{eq:sde-2}.

To prove this claim,  it is sufficient to apply the Ito-jump stochastic calculus \cite{applebaum,cont-tankov}.

In fact we see that the SDE Eq.~\eqref{eq:sde-angle-1-1} is not well defined when $\theta_t^\gamma = \pi$, so that some extra boundary conditions should be added to defined it correctly. This is due to the fact that the change of variable Eq.~\eqref{eq:CV} is  not smooth.  However, in the particular case we consider, where $\varphi_t^\gamma= \pi/2$, we can disregard this technical issue.

\subsection{Proof of Eq.~\eqref{eq:Eqn-r}}

We have $(r_{t}^{\gamma})^2=\left(2q_{t}^{\gamma}-1\right)^{2}+4u_{t}^{\gamma}{\overline u}_t^{\gamma}$
and then with Ito-jump stochastic calculus

\begin{align*}
	d (r_{t}^{\gamma})^2 & =d\left[\left(2q_{t}^{\gamma}-1\right)^{2}+4u_{t}^{\gamma}{\overline u}_t^{\gamma}\right]\\
	& =4\left(2q_{t}^{\gamma}-1\right)\left(dq_{t}^{\gamma}\right)_{c}+4\left(du_{t}^{\gamma}\right)_{c}{\overline u}_t^{\gamma}+4u_{t}^{\gamma}\left(d{\overline u}_t^{\gamma}\right)_{c}\\
	& +\left(2\left(q_{t}^{\gamma}-q_{t}^{\gamma}d{\mathcal N}_{t}^{\gamma}\right)-1\right)^{2}-\left(2q_{t}^{\gamma}-1\right)^{2}\\
	& +4\left(\left(u_{t}^{\gamma}-q_{t}^{\gamma}d{\mathcal N}_{t}^{\gamma}\right)\left({\overline u}_t^{\gamma}-q_{t}^{\gamma}d{\mathcal N}_{t}^{\gamma}\right)-u_{t}^{\gamma}{\overline u}_t^{\gamma}\right)
\end{align*}
with
\[
\begin{cases}
	\left(dq_{t}^{\gamma}\right)_{c}=ik_{\gamma}\left(u_{t}^{\gamma}-{\overline u}_t^{\gamma}\right)+\gamma\eta q_{t}^{\gamma}\left(1-q_{t}^{\gamma}\right) \ ,\\
	\left(du_{t}^{\gamma}\right)_{c}=-ik_{\gamma}\left(2q_{t}^{\gamma}-1\right)-\frac{\gamma}{2}u_{t}^{\gamma}+\gamma\eta u_{t}^{\gamma}\left(1-q_{t}^{\gamma}\right) \ , \\
	\left(d{\overline u}_t^{\gamma}\right)_{c}=+ik_{\gamma}\left(2q_{t}^{\gamma}-1\right)-\frac{\gamma}{2}{\overline u}_t^{\gamma}+\gamma\eta{\overline u}_t^{\gamma}\left(1-q_{t}^{\gamma}\right) \ .
\end{cases}
\]
We have
\begin{align*}
	&d (r_{t}^{\gamma})^2  = d\left(\left(2q_{t}^{\gamma}-1\right)^{2}+4u_{t}^{\gamma}{\overline u}_t^{\gamma}\right)\\
	& = 4\left(2q_{t}^{\gamma}-1\right)dq_{t}^{\gamma}+4 \left(dq_{t}^{\gamma}\right)\left(dq_{t}^{\gamma}\right)+4\left(du_{t}^{\gamma}\right){\overline u}_t^{\gamma}\\
	& + 4u_{t}^{\gamma}\left(d{\overline u}_t^{\gamma}\right)+4\left(du_{t}^{\gamma}\right)\left(d{\overline u}_t^{\gamma}\right)\\
	& = 4\left(2q_{t}^{\gamma}-1\right)\left(\left[-ik_{\gamma}\left(u_{t}^{\gamma}-{\overline u}_t^{\gamma}\right)+\gamma\eta\left(1-q_{t}^{\gamma}\right)q_{t}^{\gamma}\right]dt \right.\\
	& \left. \quad \quad \quad \quad \quad   -q_{t}^{\gamma}d{\mathcal N}_{t}^{\gamma}\right)\\
	& +4(q_{t}^{\gamma})^{2}d{\mathcal N}_{t}^{\gamma}\\
	&+4\left(\left[-ik_{\gamma}\left(2q_{t}^{\gamma}-1\right){\overline u}_t^{\gamma}+\gamma\left(-\tfrac{1}{2}+\eta-\eta q_{t}^{\gamma}\right)u_{t}^{\gamma}\overline{u}_t^{\gamma}\right]dt \right. \\
	& \quad \quad \quad  \left. -u_{t}^{\gamma}{\overline u}_t^{\gamma}d{\mathcal N}_{t}^{\gamma}\right)\\
	&=\gamma\left(\eta-1\right)2(1-q_{t}^{\gamma})\left((r_{t}^{\gamma})^2+2q_{t}^{\gamma}-1\right)\\
	&+\left(1-(r_{t}^{\gamma})^2 \right)\left(\gamma\left(2q_{t}^{\gamma}-1\right)dt+d{\mathcal N}_{t}^{\gamma}\right).
\end{align*}

\section{Convergence of spiking process for the scaling $k_{\gamma}\sim\gamma^\alpha$}
\label{app:Rabi-scaling}
\begin{figure*}
	\centering
	\includegraphics[width=0.42\linewidth]{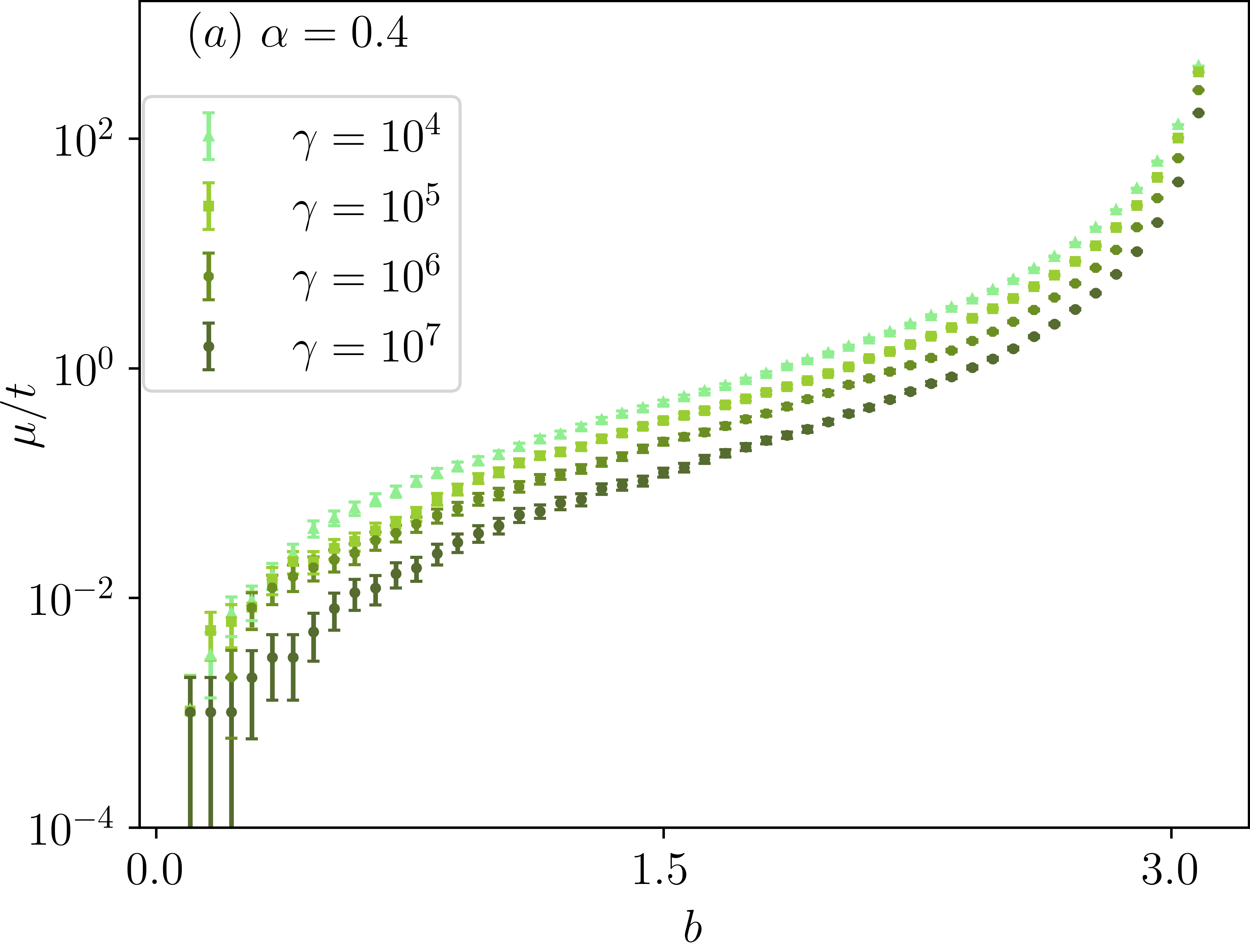}\hspace{1cm}
	\includegraphics[width=0.42\linewidth]{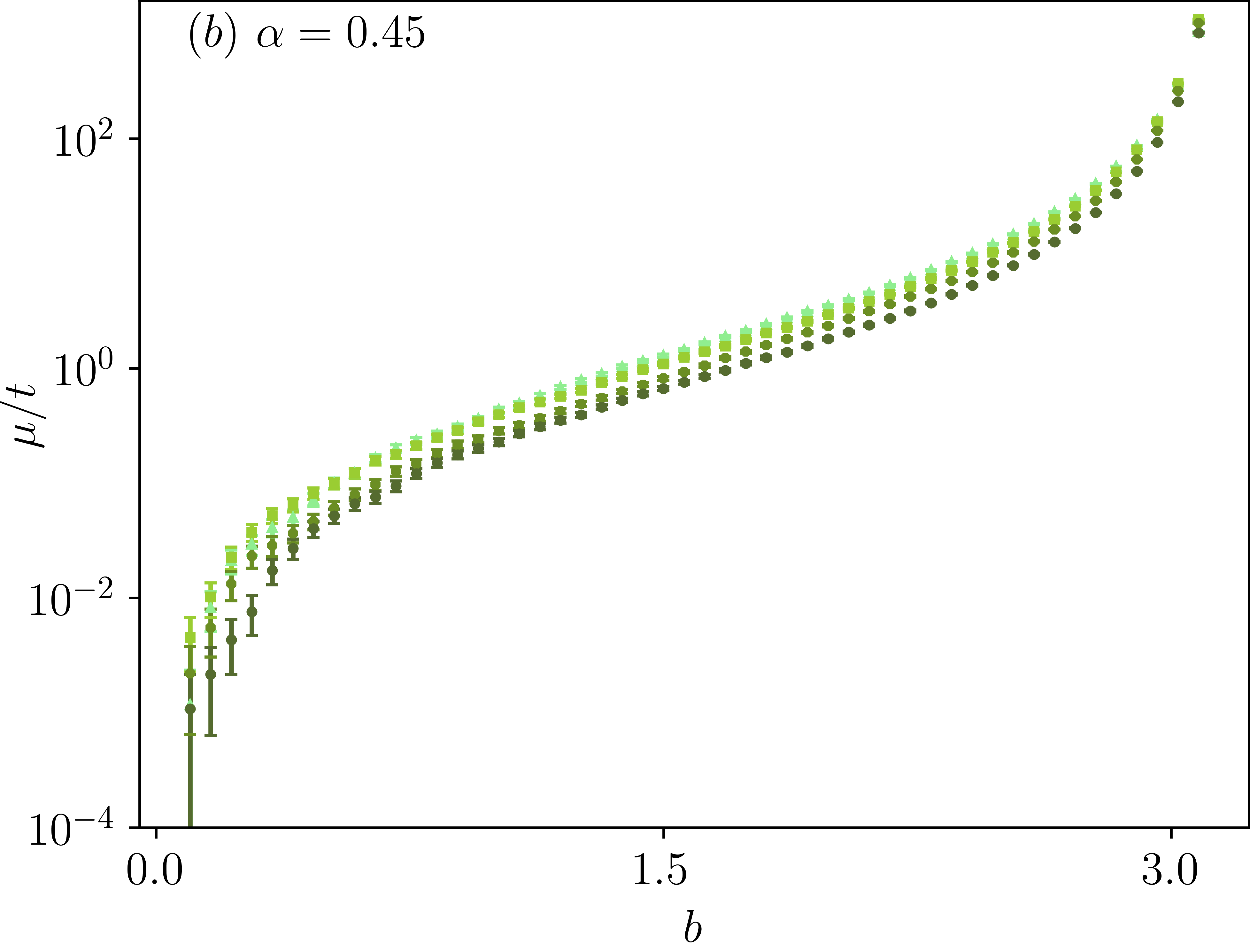}\\
	\includegraphics[width=0.42\linewidth]{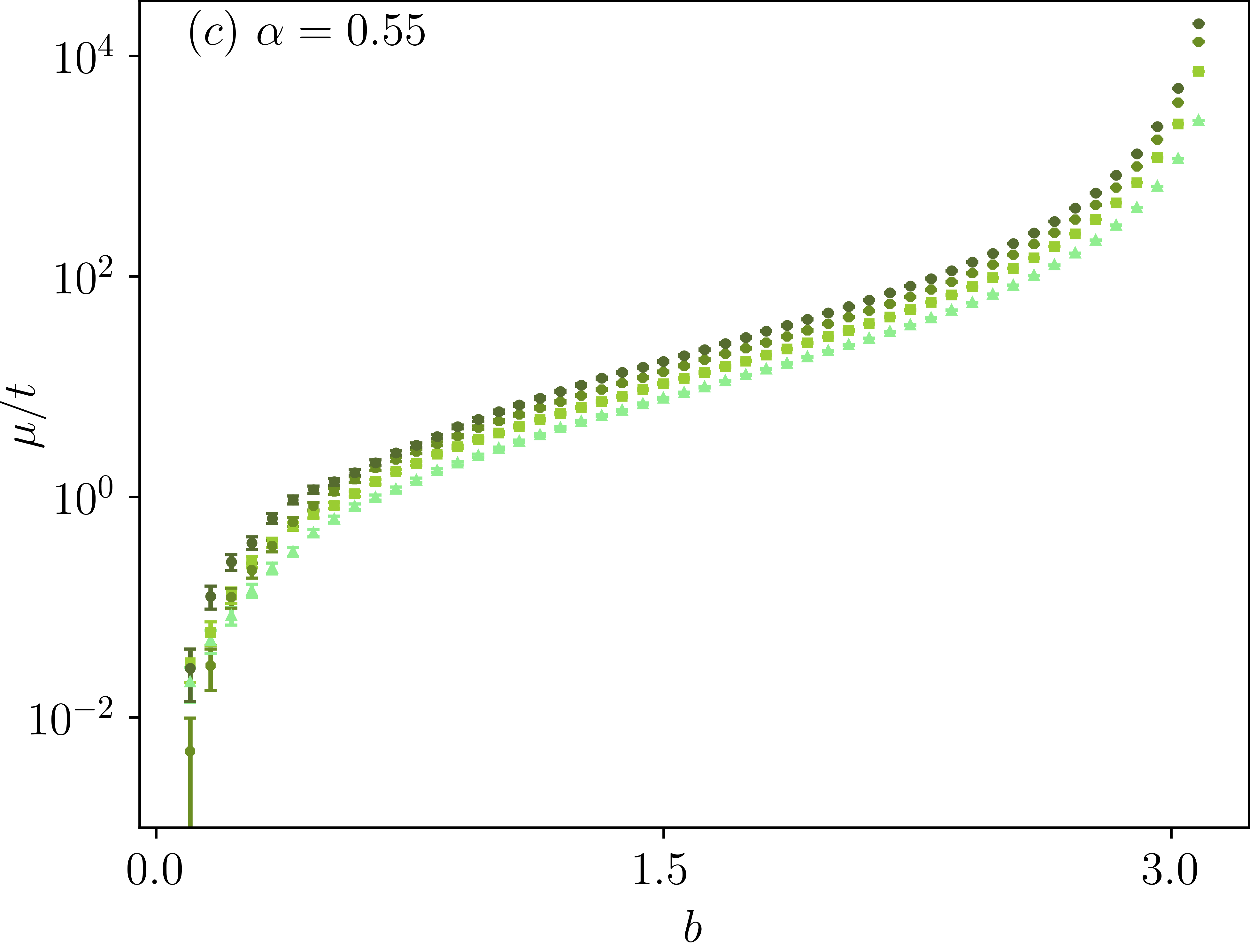}\hspace{1cm}
	\includegraphics[width=0.42\linewidth]{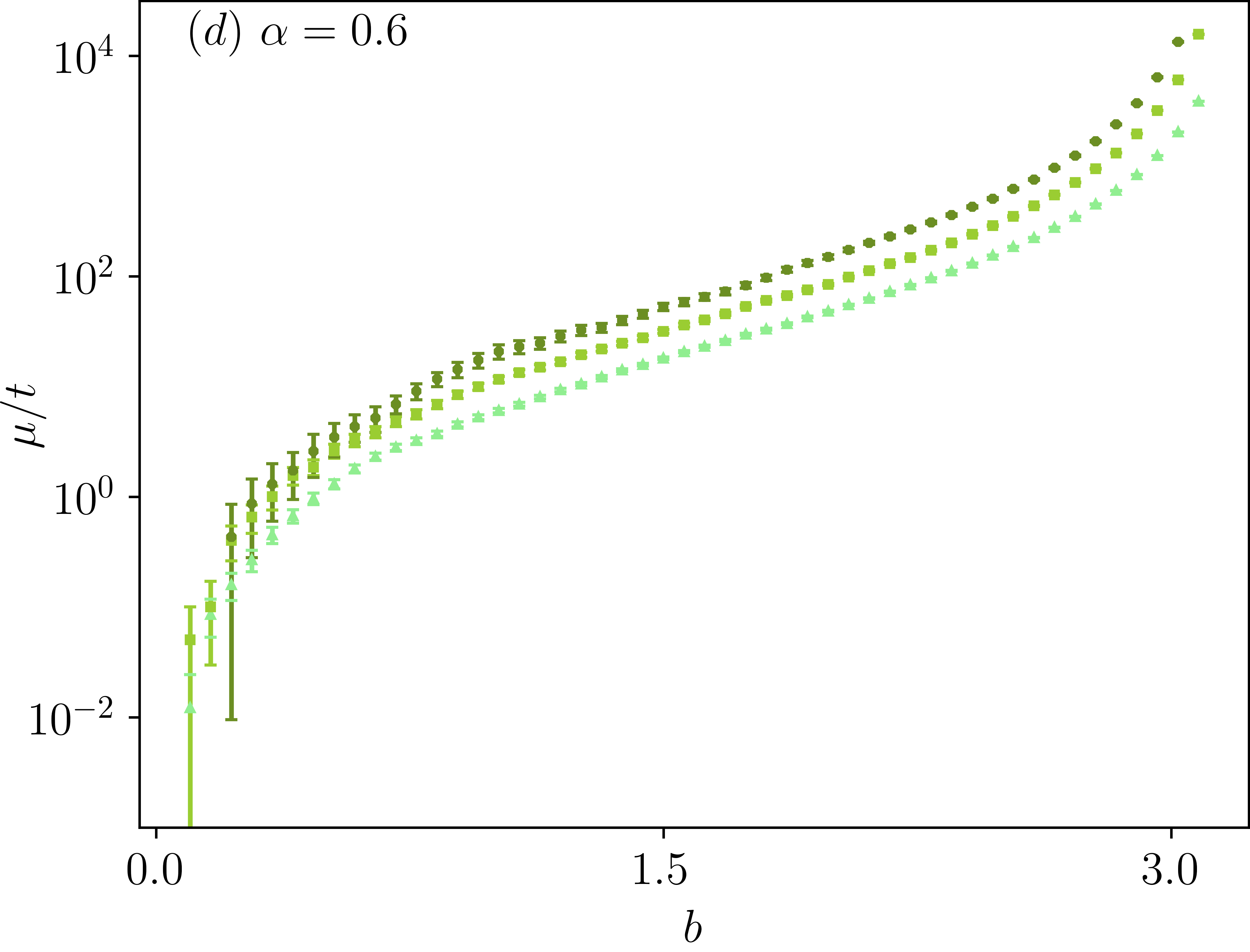}
	\caption{Mean of the distribution (conditioned on no-jumps) of the number of spikes (from $\theta_t=\pi$) per unit time in an space-time box $(0,b)\times(0,t)$ plotted against the box edge $b$ for (a) $k_\gamma\sim\gamma^{0.4}$ (b) $k_\gamma\sim\gamma^{0.45}$ (c) $k_\gamma\sim\gamma^{0.55}$ (d) $k_\gamma\sim\gamma^{0.6}$. The separation between the scatter plots for various $\gamma$ is more than the width of the error bars and clearly indicates a downward trend with increasing $\gamma$ for $\alpha=0.4$ and $\alpha=0.45$ (indicating that the spikes vanish for $\gamma\rightarrow\infty$) and clearly indicates an upward trend with increasing $\gamma$ for $\alpha=0.55$ and $\alpha=0.6$ (indicating that the moments of the spike distribution diverge for $\gamma\rightarrow\infty$). The data-points were obtained by averaging over $10^4$ realizations. Note that for $\alpha=0.6$, we were not able to obtain data for $\gamma=10^7$ since almost all the trajectories in the sample had the occurrence of jumps (recall that the spike distribution is conditioned on no-jumps).}
	\label{plts:Rabi-scaling}
\end{figure*}

We discuss the motivation behind the choice of $k_{\gamma}=\sqrt{\omega\gamma}$ in the Collapse-Unitary case in the main text. From numerical simulations (see Fig. \ref{plts:Rabi-scaling}), we determine that for the scaling $k_{\gamma}\sim \gamma^\alpha$, the distribution of the spiking process does not converge for large $\gamma$ for $\alpha\gtrapprox0.5$ and the spikes vanish for large $\gamma$ for $\alpha\lessapprox0.5$. This can also be seen by taking the limits $\omega\rightarrow0$ and $\omega\rightarrow\infty$ in the first line of Eq.~\eqref{eq:spikeintense} (corresponding to $\alpha<0.5$ and $\alpha>0.5$ respectively). This provides credible justification for the form of $k_{\gamma}$ that was assumed in the main text.

\section{Exact computation and asymptotics of Laplace transforms}
\label{app:LaplaceTransform}

\subsection{Collapse-Unitary setup}
We derive the explicit expressions for the functions $C(\sigma)$, $D(\sigma)$ and $E(\sigma)$ (and their asymptotic forms for large $\gamma$) to determine the generating function of the distribution of spikes for the Collapse-Unitary case. Thanks to the explicit expressions Eq.~\eqref{eq:mualpha-unit}, we have then that

\begin{equation}
	\begin{split}
		& E(\sigma)= \cfrac{1}{4(\beta^\prime)^2} \left\{ \cfrac{1+e^{-2\phi}}{\sigma+\gamma/2- 2\beta k_\gamma} \left(1-e^{-(\sigma+\gamma/2 -2\beta k_\gamma )\tau}\right)\right. \\
		& \quad \quad +  \cfrac{1+e^{2\phi}}{\sigma+\gamma/2+ 2\beta k_\gamma} \left(1-e^{-(\sigma+\gamma/2 +2\beta k_\gamma )\tau}\right)\\
		&\quad \quad - \left. \cfrac{4}{\sigma+\gamma/2} \left( 1-e^{-(\sigma+\gamma/2) \tau}\right) \right\} \ ,
	\end{split}
	\label{eq:E}
\end{equation}
\begin{equation}
	\begin{split}
		&C(\sigma)+D(\sigma) \\
		&= \cfrac{\gamma}{4(\beta^\prime)^2} \left\{ \cfrac{e^{-2\phi}}{\sigma+\gamma/2- 2\beta k_\gamma} \left(1 - e^{-(\sigma+\gamma/2 -2\beta k_\gamma )\tau}\right)\right. \\
		& \quad +  \cfrac{e^{2\phi}}{\sigma+\gamma/2+ 2\beta k_\gamma} \left(1 - e^{-(\sigma+\gamma/2 +2\beta k_\gamma )\tau} \right)\\
		&\quad - \left. \cfrac{2}{\sigma+\gamma/2} \left( 1 - e^{-(\sigma+\gamma/2) \tau}\right) \right\} \  ,
	\end{split}
\end{equation}
and
\begin{equation}
	\begin{split}
		& D(\sigma)\\
		&= \left\{ \tfrac{e^{-2\phi}}{\sigma+\gamma/2- 2\beta k_\gamma} \left(e^{-(\sigma+\gamma/2 -2\beta k_\gamma )\tau_b} - e^{-(\sigma+\gamma/2 -2\beta k_\gamma )\tau_a}\right)\right. \\
		&+  \tfrac{e^{2\phi}}{\sigma+\gamma/2+ 2\beta k_\gamma} \left(e^{-(\sigma+\gamma/2 +2\beta k_\gamma )\tau_b} - e^{-(\sigma+\gamma/2 +2\beta k_\gamma )\tau_a} \right)\\
		&- \left. \tfrac{2}{\sigma+\gamma/2} \left( e^{-(\sigma+\gamma/2) \tau_b} - e^{-(\sigma+\gamma/2) \tau_a}\right) \right\} \times \cfrac{\gamma}{4(\beta^\prime)^2} \ . 
	\end{split}
	\label{eq:D}
\end{equation}

We note the asymptotic forms for large $\gamma$ (here we take $k_\gamma=\sqrt{\omega\gamma}$),
\begin{equation}
	\begin{split}
		C(\sigma) & \sim 1 - (4\omega+\sigma) \, {\gamma}^{-1} -D(\sigma) \ , \\
		D(\sigma) & \sim 4\omega \gamma^{-1} \left( \tan^2 (b/2) - \tan^2 (a/2) \right) \ , \\
		E (\sigma) & \sim \gamma^{-1} \ .
	\end{split}
\end{equation}
Then Eq.~\eqref{eq:Z_Rabi} follows.

\subsection{Collapse-Thermal setup}

We derive the explicit expressions for the functions $C(\sigma)$, $D(\sigma)$ and $E(\sigma)$ (and their asymptotic forms for large $\gamma$) to determine the generating function of the distribution of spikes for the Collapse-Thermal case. Thanks to Eq.~\eqref{eq:Alan007} we have that
\begin{equation}    
	\label{eq:E-th}
	\begin{split}
		&E(\sigma)\\
		&= \tfrac{1}{q_+-q_-} \left\{ \tfrac{q_+(1-e^{-[\sigma+\eta\gamma(1-q_-)]\tau})}{\sigma+\eta\gamma(1-q_-)} \right. - \left. \tfrac{q_-(1-e^{-[\sigma+\eta\gamma(1-q_+)]\tau})}{\sigma+\eta\gamma(1-q_+)} \right\} \ , 
	\end{split}
\end{equation}

\begin{equation} 
	\label{eq:A}
	\begin{split}
		C(\sigma)+D(\sigma)= \tfrac{\gamma\eta}{q_+-q_-} &\left\{ \tfrac{q_+(1-q_-)(1-e^{-[\sigma+\eta\gamma(1-q_-)]\tau})}{\sigma+\eta\gamma(1-q_-)} \right.  \\
		& \quad - \left. \tfrac{q_-(1-q_+)(1-e^{-[\sigma+\eta\gamma(1-q_+)]\tau})}{\sigma+\eta\gamma(1-q_+)} \right\} \ , 
	\end{split} 
\end{equation}
and 
\begin{equation} 
	\label{eq:D-th}
	\begin{split}
		& D(\sigma)= \tfrac{\gamma\eta}{q_+-q_-} \left\{ \tfrac{q_+(1-q_-)(e^{-[\sigma+\eta\gamma(1-q_-)]\tau_a}-e^{-[\sigma+\eta\gamma(1-q_-)]\tau_b})}{\sigma+\eta\gamma(1-q_-)} \right. \\
		&\quad  \quad - \left. \tfrac{q_-(1-q_+)(e^{-[\sigma+\eta\gamma(1-q_+)]\tau_a}-e^{-[\sigma+\eta\gamma(1-q_+)]\tau_b})}{\sigma+\eta\gamma(1-q_+)} \right\} \ .
	\end{split}
\end{equation}
where we have taken the limit $\epsilon\rightarrow 0$. We note then that the asymptotic forms of $E(\sigma),C(\sigma)$ and $D(\sigma)$ for large $\gamma$ are given by
\begin{equation}
	\label{asym}
	\begin{split}
		E(\sigma)& \sim\frac{1}{\sigma+\gamma\eta+W_{-,+}} \ , \\
		C(\sigma) & \sim \frac{\gamma\eta-W_{-,+}\Big(\frac{1}{a}-\frac{1}{b}\Big)}{\sigma+\gamma\eta+W_{-,+}} \ , \\
		D(\sigma)& \sim \frac{W_{-,+}\Big(\frac{1}{a}-\frac{1}{b}\Big)}{\sigma+\gamma\eta+W_{-,+}} \ .
	\end{split}
\end{equation}
Then Eq.~\eqref{eq:Z_thm} follows.

\subsection{Collapse-Measurement setup}
We derive the explicit expressions for the functions $C(\sigma)$, $D(\sigma)$ and $E(\sigma)$ (and their asymptotic forms for large $\gamma$) to determine the generating function of the distribution of spikes for the Collapse-Measurement case. Thanks to Eq.~\eqref{eq:Alan014} we have that
\begin{equation}    
	\label{eq:E-cc}
	\begin{split}
		E(\sigma)= \frac{1}{\gamma_1\eta_1+\gamma_2} & \left\{\tfrac{\gamma_2(1-\eta_2)(1-e^{-\sigma\tau})}{\sigma} \right.\\
		& \left. +\tfrac{(\gamma_1\eta_1+\gamma_2\eta_2)(1-e^{-(\sigma+\gamma_2+\gamma_1\eta_1)\tau})}{\sigma+\gamma_2+\gamma_1\eta_1}\right\} \ , 
	\end{split}
\end{equation}
\begin{equation}
	\label{eq:A-cc}
	\begin{split}
		C(\sigma)+D(\sigma)=\frac{\gamma_1\eta_1(1-e^{-(\sigma+\gamma_2+\gamma_1\eta_1)\tau})}{\sigma+\gamma_1\eta_1+\gamma_2} \ ,
	\end{split}
\end{equation}
and 
\begin{equation}
	\begin{split}
		D(\sigma)= \frac{\gamma_1\eta_1(e^{-(\sigma+\gamma_1\eta_1+\gamma_2)\tau_a}-e^{-(\sigma+\gamma_1\eta_1+\gamma_2)\tau_b})}{\sigma+\gamma_1\eta_1+\gamma_2} \ .
	\end{split}
	\label{eq:D-cc}
\end{equation}
We note then that the asymptotic forms of $E(\sigma),C(\sigma)$ and $D(\sigma)$ for large $\gamma_1$ are given by
\begin{equation}
	\begin{split}
		E(\sigma)&\sim\frac{1-\frac{\epsilon\gamma_2(1-\eta_2)}{\sigma+\gamma_1\eta_1+\gamma_2}}{\sigma+\gamma_1\eta_1+\gamma_2} \ ,  \\
		C(\sigma)&\sim \frac{\gamma_1\eta_1-\gamma_2(1-\eta_2)\Big(\epsilon+\frac{1}{a}-\frac{1}{b}\Big)}{\sigma+\gamma_1\eta_1+\gamma_2} \ , \\
		D(\sigma)&\sim\frac{\gamma_2(1-\eta_2)\Big(\frac{1}{a}-\frac{1}{b}\Big)}{\sigma+\gamma_1\eta_1+\gamma_2} \ .
	\end{split}
\end{equation}
Then Eq.~\eqref{eq:Z-cc} follows by taking the limit $\gamma_1\rightarrow\infty$ and $\epsilon\rightarrow 0$, in that order.

\section{Numerical Methods}
Here, we discuss the two numerical methods employed to simulate the 1D stochastic differential equations (SDEs) with Poissonian noise that govern the evolution of the state $x_t$. The first method is a first-order iterative approach where, given an SDE with Poissonian noise:

\begin{equation}
	\begin{split}
		dx_t&=f(x_t)dt+h(x_t)d{\mathcal N}_t \ ,\\
		d{\mathcal N}_td{\mathcal N}_t&=d{\mathcal N}_t\ ,\\
		\mathbb{E}[d{\mathcal N}_t]&=g(x_t)dt\ ,
	\end{split}
	\label{eq:PoissSDE}
\end{equation}

$\{x_t\}_{t=\Delta t,2\Delta t...}$ is obtained using the rule
\begin{equation}
	x_{t+\Delta t} = x_t + f(x_t)\Delta t + h(x_t)r_t\ ,
\end{equation}
where the step-size $\Delta t\ll1$ is chosen appropriately and the random number $r_t$ is sampled from a Bernoulli distribution with the parameter $p=g(x_t)\Delta t$.
\\\\
The second method is based on the principle that the stochastic evolution Eq.~\eqref{eq:PoissSDE} is composed of deterministic flows and jumps (not to be confused with quantum jumps). Discarding the Poissonian noise, Eq.~\eqref{eq:PoissSDE} becomes a first order ODE: $\dot{x}_t=f(x_t)$ which, given the initial condition $x_{t=0}=x_0$, has the unique solution $x_{t,x_0}$. Now, the probability $\nu(t,x_0)$ of having no-jumps in an interval $(0,t)$ with the initial condition $x_{t=0}=x_0$ is formally given by:
\begin{equation}
	\nu(t,x_0)=\exp\Bigg(-\int_{x_0}^{x_{t,x_0}}\tfrac{g(x)dx}{f(x)}\Bigg)\ ,
	\label{eq:no-jump-pro}
\end{equation}

We now have all the ingredients necessary for this method. Given that the state at time $t_i$ is $x_{t_i}$, the algorithm for time evolution is as follows:
\begin{enumerate}
	\item Choose a random number $\tau$ from the distribution $-\tfrac{\partial\nu(t,x_{t_i})}{\partial t}$, where $\nu(t,x_{t_i})$ is given by Eq.~\eqref{eq:no-jump-pro}. Physically speaking, this means that a jump has occurred at time $t_{i+1}=t_i+\tau$ and from the deterministic flow, we know the state takes the values $x_{t'}=x_{t'-t_i,x_{t_i}}$ for $t_i\leq t'< t_{i+1}$.
	\item The value of the state at time $t_{i+1}$ is set by $x_{t_{i+1}}=x_{\tau,x_{t_i}}+h(x_{\tau,x_{t_i}})$. \item Repeat steps 1 \& 2 with the transformations $t_i\rightarrow t_{i+1}$ and $t_{i+1}\rightarrow t_{i+2}$.
\end{enumerate}

\bibliography{spikes}

\end{document}